\newif\ifprd
\prdtrue

\ifprd
\documentclass[aps,prd,twocolumn,superscriptaddress,showpacs]{revtex4}

\usepackage{graphicx}
\usepackage{dcolumn}
\usepackage{bm}

\else
\RequirePackage{lineno}
\documentclass[prd,amsmath,amssymb,preprint]{revtex4}
\usepackage{graphicx}
\usepackage{dcolumn}
\usepackage{bm}
\linenumbers
\pagewiselinenumbers

\linenumbers

\fi

\begin{document}
\preprint{CDF/PUB/JET/PUBLIC/8996}
\preprint{PRD Draft Version 2.0}

\title{Measurement of $b$-jet Shapes in Inclusive Jet Production in $p\bar{p}$ Collisions at $\sqrt{s} = 1.96$ TeV}

\date{\today}

\begin{abstract}

We present a measurement of the shapes of $b$-jets using 300~$\mbox{pb}^{-1}$ of data obtained with the upgraded Collider Detector at Fermilab (CDF II) in $p\bar p$ collisions at center of mass energy $\sqrt{s}=1.96$ TeV. This measurement covers a wide transverse momentum range, from 52 to 300~GeV/$c$. Samples of heavy-flavor enhanced jets together with inclusive jets are used to extract the average shapes of $b$-jets. The $b$-jets are expected to be broader than inclusive jets. Moreover, $b$-jets containing a single $b$-quark are expected to be narrower than those containing a $b\bar{b}$ pair from gluon splitting. The measured $b$-jet shapes are found to be significantly broader than expected from the {\sc pythia} and {\sc herwig} Monte Carlo simulations.  This effect may arise from an underestimation of the fraction of $b$-jets originating from gluon splitting in these simulations. The jet shape distributions provided in this paper could be compared to any full Monte Carlo simulation and could be used to further constrain the various parameters.

\end{abstract}


\ifprd
\affiliation{Institute of Physics, Academia Sinica, Taipei, Taiwan 11529, Republic of China} 
\affiliation{Argonne National Laboratory, Argonne, Illinois 60439} 
\affiliation{University of Athens, 157 71 Athens, Greece} 
\affiliation{Institut de Fisica d'Altes Energies, Universitat Autonoma de Barcelona, E-08193, Bellaterra (Barcelona), Spain} 
\affiliation{Baylor University, Waco, Texas  76798} 
\affiliation{Istituto Nazionale di Fisica Nucleare Bologna, $^t$University of Bologna, I-40127 Bologna, Italy} 
\affiliation{Brandeis University, Waltham, Massachusetts 02254} 
\affiliation{University of California, Davis, Davis, California  95616} 
\affiliation{University of California, Los Angeles, Los Angeles, California  90024} 
\affiliation{University of California, San Diego, La Jolla, California  92093} 
\affiliation{University of California, Santa Barbara, Santa Barbara, California 93106} 
\affiliation{Instituto de Fisica de Cantabria, CSIC-University of Cantabria, 39005 Santander, Spain} 
\affiliation{Carnegie Mellon University, Pittsburgh, PA  15213} 
\affiliation{Enrico Fermi Institute, University of Chicago, Chicago, Illinois 60637} 
\affiliation{Comenius University, 842 48 Bratislava, Slovakia; Institute of Experimental Physics, 040 01 Kosice, Slovakia} 
\affiliation{Joint Institute for Nuclear Research, RU-141980 Dubna, Russia} 
\affiliation{Duke University, Durham, North Carolina  27708} 
\affiliation{Fermi National Accelerator Laboratory, Batavia, Illinois 60510} 
\affiliation{University of Florida, Gainesville, Florida  32611} 
\affiliation{Laboratori Nazionali di Frascati, Istituto Nazionale di Fisica Nucleare, I-00044 Frascati, Italy} 
\affiliation{University of Geneva, CH-1211 Geneva 4, Switzerland} 
\affiliation{Glasgow University, Glasgow G12 8QQ, United Kingdom} 
\affiliation{Harvard University, Cambridge, Massachusetts 02138} 
\affiliation{Division of High Energy Physics, Department of Physics, University of Helsinki and Helsinki Institute of Physics, FIN-00014, Helsinki, Finland} 
\affiliation{University of Illinois, Urbana, Illinois 61801} 
\affiliation{The Johns Hopkins University, Baltimore, Maryland 21218} 
\affiliation{Institut f\"{u}r Experimentelle Kernphysik, Universit\"{a}t Karlsruhe, 76128 Karlsruhe, Germany} 
\affiliation{Center for High Energy Physics: Kyungpook National University, Daegu 702-701, Korea; Seoul National University, Seoul 151-742, Korea; Sungkyunkwan University, Suwon 440-746, Korea; Korea Institute of Science and Technology Information, Daejeon, 305-806, Korea; Chonnam National University, Gwangju, 500-757, Korea} 
\affiliation{Ernest Orlando Lawrence Berkeley National Laboratory, Berkeley, California 94720} 
\affiliation{University of Liverpool, Liverpool L69 7ZE, United Kingdom} 
\affiliation{University College London, London WC1E 6BT, United Kingdom} 
\affiliation{Centro de Investigaciones Energeticas Medioambientales y Tecnologicas, E-28040 Madrid, Spain} 
\affiliation{Massachusetts Institute of Technology, Cambridge, Massachusetts  02139} 
\affiliation{Institute of Particle Physics: McGill University, Montr\'{e}al, Canada H3A~2T8; and University of Toronto, Toronto, Canada M5S~1A7} 
\affiliation{University of Michigan, Ann Arbor, Michigan 48109} 
\affiliation{Michigan State University, East Lansing, Michigan  48824}
\affiliation{Institution for Theoretical and Experimental Physics, ITEP, Moscow 117259, Russia} 
\affiliation{University of New Mexico, Albuquerque, New Mexico 87131} 
\affiliation{Northwestern University, Evanston, Illinois  60208} 
\affiliation{The Ohio State University, Columbus, Ohio  43210} 
\affiliation{Okayama University, Okayama 700-8530, Japan} 
\affiliation{Osaka City University, Osaka 588, Japan} 
\affiliation{University of Oxford, Oxford OX1 3RH, United Kingdom} 
\affiliation{Istituto Nazionale di Fisica Nucleare, Sezione di Padova-Trento, $^u$University of Padova, I-35131 Padova, Italy} 
\affiliation{LPNHE, Universite Pierre et Marie Curie/IN2P3-CNRS, UMR7585, Paris, F-75252 France} 
\affiliation{University of Pennsylvania, Philadelphia, Pennsylvania 19104} 
\affiliation{Istituto Nazionale di Fisica Nucleare Pisa, $^q$University of Pisa, $^r$University of Siena and $^s$Scuola Normale Superiore, I-56127 Pisa, Italy} 
\affiliation{University of Pittsburgh, Pittsburgh, Pennsylvania 15260} 
\affiliation{Purdue University, West Lafayette, Indiana 47907} 
\affiliation{University of Rochester, Rochester, New York 14627} 
\affiliation{The Rockefeller University, New York, New York 10021} 

\affiliation{Istituto Nazionale di Fisica Nucleare, Sezione di Roma 1, $^v$Sapienza Universit\`{a} di Roma, I-00185 Roma, Italy} 

\affiliation{Rutgers University, Piscataway, New Jersey 08855} 
\affiliation{Texas A\&M University, College Station, Texas 77843} 
\affiliation{Istituto Nazionale di Fisica Nucleare Trieste/\ Udine, $^w$University of Trieste/\ Udine, Italy} 
\affiliation{University of Tsukuba, Tsukuba, Ibaraki 305, Japan} 
\affiliation{Tufts University, Medford, Massachusetts 02155} 
\affiliation{Waseda University, Tokyo 169, Japan} 
\affiliation{Wayne State University, Detroit, Michigan  48201} 
\affiliation{University of Wisconsin, Madison, Wisconsin 53706} 
\affiliation{Yale University, New Haven, Connecticut 06520} 
\author{T.~Aaltonen}
\affiliation{Division of High Energy Physics, Department of Physics, University of Helsinki and Helsinki Institute of Physics, FIN-00014, Helsinki, Finland}
\author{J.~Adelman}
\affiliation{Enrico Fermi Institute, University of Chicago, Chicago, Illinois 60637}
\author{T.~Akimoto}
\affiliation{University of Tsukuba, Tsukuba, Ibaraki 305, Japan}
\author{M.G.~Albrow}
\affiliation{Fermi National Accelerator Laboratory, Batavia, Illinois 60510}
\author{B.~\'{A}lvarez~Gonz\'{a}lez}
\affiliation{Instituto de Fisica de Cantabria, CSIC-University of Cantabria, 39005 Santander, Spain}
\author{S.~Amerio$^u$}
\affiliation{Istituto Nazionale di Fisica Nucleare, Sezione di Padova-Trento, $^u$University of Padova, I-35131 Padova, Italy} 

\author{D.~Amidei}
\affiliation{University of Michigan, Ann Arbor, Michigan 48109}
\author{A.~Anastassov}
\affiliation{Northwestern University, Evanston, Illinois  60208}
\author{A.~Annovi}
\affiliation{Laboratori Nazionali di Frascati, Istituto Nazionale di Fisica Nucleare, I-00044 Frascati, Italy}
\author{J.~Antos}
\affiliation{Comenius University, 842 48 Bratislava, Slovakia; Institute of Experimental Physics, 040 01 Kosice, Slovakia}
% ...

\collaboration{CDF Collaboration\footnote{With visitors from $^a$Universiteit Antwerpen, B-2610 Antwerp, Belgium, 
$^b$Chinese Academy of Sciences, Beijing 100864, China, 
$^c$University of Bristol, Bristol BS8 1TL, United Kingdom, 
$^d$University of California Irvine, Irvine, CA  92697, 
$^e$University of California Santa Cruz, Santa Cruz, CA  95064, 
$^f$Cornell University, Ithaca, NY  14853, 
$^g$University of Cyprus, Nicosia CY-1678, Cyprus, 
$^h$University College Dublin, Dublin 4, Ireland, 
$^i$University of Edinburgh, Edinburgh EH9 3JZ, United Kingdom, 
$^j$Universidad Iberoamericana, Mexico D.F., Mexico, 
$^k$University of Manchester, Manchester M13 9PL, England, 
$^l$Nagasaki Institute of Applied Science, Nagasaki, Japan, 
$^m$University de Oviedo, E-33007 Oviedo, Spain, 
$^n$Queen Mary, University of London, London, E1 4NS, England, 
$^o$Texas Tech University, Lubbock, TX  79409, 
$^p$IFIC(CSIC-Universitat de Valencia), 46071 Valencia, Spain,
$^x$Royal Society of Edinburgh, 
}}
\noaffiliation

\fi

\maketitle

\section{Introduction}\label{sec:Intro}
The measurement of jet shapes allows a study of the processes that occur between the initial hard interaction and the collimated flow of hadrons observed experimentally~\cite{general}. The internal structure of jets is dictated principally by the multiple gluon emissions from the primary parton. Multi-gluon emission involves high order QCD processes that are complicated to calculate, and various parton shower models are used to implement this in Monte Carlo (MC) simulation. In addition to this, a good understanding of the hadronization and fragmentation processes is needed in order to compare simulation results with what is observed in data. For heavy flavor jets, the decay of the heavy hadrons must also be correctly modeled. Jet shapes, that describe the transverse momentum distribution inside jets as a function of distance from the jet axis, are defined in detail in Sec.~\ref{sec:jet_shapes_def}. They are interesting distributions to measure in order to study the overall decay structure leading to the observed jets. Moreover, the underlying event, an important component of any hadronic collision, contributes to the overall jet shapes. The underlying event comprises initial- and final-state radiation, multiple parton interactions, and beam-beam remnants~\cite{pythia_tune_a_better}.\\
In this paper, the jet shapes for $b$-quark jets are measured using the data collected with the upgraded Collider Detector at Fermilab (CDF II)~\cite{cdf}. The results are compared to two different leading order Monte Carlo models; the first model is a tuned version of {\sc pythia}~\cite{pythia}, the so-called {\sc pythia} Tune A~\cite{pythia_tune_a_better}~\cite{pythia_tune_a}, and the second is {\sc herwig}~\cite{herwig}. These models are described in Sec.~\ref{sec:mc}. Jet shapes are known to be sensitive to whether the initial hard-scattered parton is a quark or a gluon; it is also expected that jet shapes are sensitive to the initial quark flavor. In the case of heavy flavor jets, the shapes are expected to be sensitive to the relative contributions of the different production mechanisms. The $b$ and the $\bar{b}$ quarks from gluon splitting are expected to be most often inside the same final jet~\cite{heavy_flavor_summary}, leading to significantly broader jet shapes than for jets originating from flavor creation. The fraction of gluon splitting events is an important parameter for the tuning of Monte Carlo simulations. \\
The inclusive jet shapes have been previously measured at CDF II~\cite{shapes_incl}. The measured jet shape variables are found to be well described by {\sc pythia} Tune A MC simulation. The data were found to not agree as well with the {\sc herwig} MC simulation. \\
Charm jet shapes were studied by the H1 collaboration in photoproduction at the Hadron-Electron Ring Accelerator at DESY (HERA)~\cite{H1}. Deviations of {\sc pythia} from measured data were observed in the region where the gluon splitting to $c\bar{c}$ pair is expected, from simulation, to contribute significantly. The study found no deviations in inclusive di-jet photoproduction.\\
 Some measurements investigating correlations between $b\bar{b}$ pairs have shown indications of an enhancement over the LO expectations of the contribution from gluon splitting. The CDF collaboration~\cite{cdf_bbbar_low} measured $b\bar{b}$ azimuthal correlations in $p\bar{p}$ collisions at 1.8 TeV by requiring two jets to be heavy-flavor tagged in semileptonic $b$-hadron decay and analyzing the azimuthal angle difference $\Delta\phi$ between these two jets. 
The transverse momentum range was lower than that reported in the present paper. The minimum transverse momentum, $p_T$~\cite{foot:eta}, for $e$ or $\mu$ leptons was 8~GeV/$c$, from which a minimum $b$-hadron $p_T$ of 14~GeV/$c$ was inferred. From the enhanced yield at small $\Delta \phi$, it was concluded that the contribution of gluon splitting in heavy flavor had to be roughly doubled over expectations from the leading order {\sc pythia} Monte Carlo models. The predictions from a next to leading order Monte Carlo simulation show good agreement with the data.\\
Similar conclusions were reached by the D\O\ collaboration~\cite{d0_bbbar} in a study of $b\bar{b}$ production cross section and azimuthal correlation using single muon and dimuon samples. The $b$-hadron $p_T$ range was $6<p_T^b<30$~GeV/$c$. They used a next to leading order (NLO) Monte Carlo simulation, and found good agreement with data in the azimuthal correlation between the two muons. The NLO calculation enhances the $b\bar{b}$ production in the $\Delta \phi<1$ region substantially over the expectations of leading order, which has no gluon splitting in the matrix element. \\
A third measurement has recently been reported by CDF~\cite{cdf_bbbar} for jet transverse energies $E_T>35$~GeV~\cite{foot:eta}.  The heavy flavor content was tagged by secondary vertices, and $\Delta \phi$ was defined as the azimuthal angle between the two secondary vertices.  Like the other two studies, an enhancement was observed for $\Delta \phi<1$, above the expectations of {\sc pythia} and {\sc herwig} Monte Carlos simulations.\\
The shape of $b$-quark jets provides another method of studying heavy flavor production mechanisms. This analysis aims to investigate if the fraction of $b$-jets originating from gluon splitting, as well as its evolution with $p_T$, is well described in the Monte Carlo models. This is particularly important for extrapolations to higher energies, such as at the LHC. The comparison of measured variables related to the internal structure of $b$-jets, such as jet shapes, to Monte Carlo simulations is sensitive to the global effect of combining models for $b$-quark production, fragmentation, hadronization, and $b$-hadron decay. \\
This paper presents a measurement of the integrated $b$-jet shape variable in inclusive jet production in $p\bar{p}$ collisions at $\sqrt{s} = 1.96$~TeV. The data were obtained using the CDF II detector. The integrated luminosity of the data sample, collected from March 2002 to August 2004, is about 300~$ \mathrm{pb}^{-1}$. Jets are reconstructed using the MidPoint cone algorithm with a cone size of 0.7. Two data sets are used for this measurement: a sample of jets from inclusive jet production, containing only a negligible amount of $b$-jets and a sample where the heavy flavor fraction has been enhanced by requiring a reconstructed secondary vertex inside the jet that comes from the decay of a heavy hadron. The integrated $b$-jet shape variable for four different $p_T$ regions covering the range from 52 to 300~GeV/$c$ are extracted from these samples after correcting for the biases introduced by the tagging as well as detector effects, as described in Sec.~\ref{sec:unfolding}. These corrections are obtained from Monte Carlo samples which are passed through a full CDF detector simulation based on {\sc geant}3~\cite{geant} where the {\sc gflash}~\cite{gflash} package is used to simulate the energy deposition in the calorimeters. The results are compared to the leading order Monte Carlo predictions from {\sc pythia} Tune A and {\sc herwig}.\\
This paper is organized as follows. In Sec.~\ref{sec:cdf} we present a brief description of the most important detector subsystems for this analysis. The jet algorithm used to reconstruct the jets used in this analysis, MidPoint, is described in Sec.~\ref{sec:midpoint}. Section~\ref{sec:jet_shapes_def} defines the function used to describe the jet shapes. Section~\ref{sec:data} describes the data samples that are used for this analysis along with the event selection. A description of the different MC models used in this analysis, both in the extraction of correction factors and for comparisons of the final results, is to be found in Sec.~\ref{sec:mc}. In Sec.~\ref{sec:unfolding} the method used to extract the $b$-jet shape variable from a sample of inclusive jets and from one of heavy-flavor enhanced jets is presented. The systematic uncertainties are presented in Sec.~\ref{sec:sys}, followed by the results in Sec.~\ref{sec:results}. Conclusions are summarized in Sec.~\ref{sec:conclusions}.
\section{The CDF II Detector Overview}\label{sec:cdf}
This section presents the  CDF II detector, a general purpose detector with azimuthal and forward-backward symmetry. It is composed of independent subsystems designed for distinct tasks related to the study of $p\bar{p}$ interactions. The two most relevant systems for this analysis are the tracking detectors and the calorimeters. A complete description of the CDF II detector can be found elsewhere~\cite{cdf}.\\
The tracking system consists of a large open-cell drift chamber and silicon microstrip detectors. These lie inside a superconducting solenoid coaxial with the beam which produces a magnetic field of 1.4~T. The fiducial region of the silicon detector covers the pseudorapidity range $|\eta| \leq 2$; this subsystem, the closest to the beam-pipe and with the finest segmentation, is used for reconstructing displaced vertices. The drift chamber measures the momentum of charged particles up to $|\eta| \leq 1$.  Segmented sampling electromagnetic and hadronic calorimeters, arranged in a projective tower geometry, surround the solenoid magnet and measure the energy flow of interacting particles in the region $|\eta| \leq 3.6$. The central barrel calorimeters cover the region $|\eta| \leq 1$, the most relevant region for this paper. The segmentation in $\phi$ and $\eta$ for both the central electromagnetic and hadronic calorimeters is $15^o$ and 0.11, respectively. The transverse momentum associated with a given calorimeter tower is obtained by assuming that the total tower momentum is given by the tower energy; the direction of the momentum vector is taken to be parallel to the vector linking the primary interaction and that tower; the transverse momentum is the projection of the momentum vector onto the plane perpendicular to the beam direction. 
Finally, a three-level trigger system~\cite{trigger} is used to select events online, as described in the section on event selection.

\section{Jet Reconstruction}\label{sec:midpoint}
In this section, the reconstruction algorithm used to reconstruct the jets used in this analysis is described. 
Jets used in this analysis are reconstructed using the MidPoint cone algorithm~\cite{midpoint}~\cite{cdf_midpoint}.\\
Before any jet algorithm is run, the electromagnetic and hadronic sections of each calorimeter tower are combined into physics towers. Each section is identified with the vector joining the primary vertex of the interaction and the section's geometrical center. The four-vector momentum components, $P \equiv$ ($p_x$,$p_y$, $p_z$,$E$), of each physics tower are then computed using the four-momentum sum of its electromagnetic and hadronic components; only towers with transverse momentum above 0.1~GeV/$c$ are considered for jet reconstruction. No contributions for the mass of particles are included in this calculation.\\
The MidPoint jet algorithm is then run over these physics towers. Each physics tower with $p_T>1$~GeV/$c$ is used to define a seed around which a jet can be formed. Starting from the seed in the event with the highest $p_T$, a cone is drawn around each seed. The radius of this search cone is half the jet cone size, measured in ($y$,$\phi$)-space~\cite{foot:eta}. 
The sum of the 4-momenta of all physics towers inside the cluster defines the 4-momentum of the cluster. The rapidity, $y$ and $\phi$ of the cluster are computed as
\ifprd
\begin{eqnarray}
y^\mathrm{cl} = \frac{1}{2}\ln{(\frac{E^\mathrm{cl}+ p_z^\mathrm{cl}}{E^\mathrm{cl} - p_z^\mathrm{cl}})} \, , \, \phi^\mathrm{cl} = \tan^{-1}(\frac{p_y^\mathrm{cl}}{p_x^\mathrm{cl}}).
\end{eqnarray}
\else
\begin{linenomath*}
\begin{equation}
y^\mathrm{cl} = \frac{1}{2}\ln{(\frac{E^\mathrm{cl}+ p_z^\mathrm{cl}}{E^\mathrm{cl} - p_z^\mathrm{cl}})} \, , \, \phi^\mathrm{cl} = \tan^{-1}(\frac{p_y^\mathrm{cl}}{p_x^\mathrm{cl}}).
\end{equation}
\end{linenomath*}
\fi
Starting from these clusters, the energy-weighted centroid including all contributions from the towers within the cluster is computed. This new point is then used as the center for a new cluster. This procedure is repeated until a stable solution is found. This occurs when the geometrical centre of the cluster is aligned with its energy-weighted centroid. In the next step, the midpoint between each pair of stable cluster centroids, separated by less than twice the jet cone radius, is added to the list of cluster centroids. The clustering algorithm is again iterated until the new set of stable clusters is found. Finally the cluster size is increased from $R/2$ to the jet cone size, $R$. At this point, if two jets overlap, the momentum sharing is considered; if the fraction of the momentum of the jet that overlaps with another jet is larger than $75\%$, the jets are merged. Else the towers are associated to the jet whose center is closest. This jet algorithm can be applied in a similar way to final state hadrons in MC generated events instead of the physics towers.\\
The size of the jet cone is chosen to be 0.7 for this analysis in order to be comparable to the inclusive jet shape measurement from CDF~\cite{shapes_incl}. A larger jet cone will result in a larger fraction of gluon jets where the gluon splits into a $q\bar{q}$ pair and both quarks end up inside the same jet. A smaller jet cone might result in this topology being reconstructed as two jets, depending on the amount of parton showering.

\section{Jet Shape Variable Definition}\label{sec:jet_shapes_def}
This section presents the jet shape variable used in this analysis. Jet shapes are defined as the distribution, as a function of the distance away from the jet axis, of the fractional transverse momentum inside the jet; jet shapes therefore measure the fraction of the total jet transverse momentum inside a given radius in ($y$,$\phi$)-space from the jet axis. The integrated jet shape variable, $\Psi(r/R)$, is defined as the fraction of total $p_T$ inside the jet cone of radius $R$ that is carried by the particles in a sub-cone of radius $r$. This quantity can be computed at the hadron level, using all final state hadrons, in the Monte Carlo simulations and by using calorimeter energy deposits or charged particle tracks for both the data and the full simulation. The measured quantity is the average integrated jet shape, which is computed over an ensemble of jets. This quantity is expressed as\\
\ifprd
\begin{eqnarray}
\Psi(r/R) = \left< \frac{p_T(0 \rightarrow r)}{p_T(0 \rightarrow R)} \right>\label{equ:int_shape_def},
\end{eqnarray}
\else
\begin{linenomath*}
\begin{equation}
\Psi(r/R) = \left< \frac{p_T(0 \rightarrow r)}{p_T(0 \rightarrow R)} \right>\label{equ:int_shape_def},
\end{equation}
\end{linenomath*}
\fi
where $p_T(0 \rightarrow r)$ is the scalar sum of the transverse momenta of all objects inside a sub-cone of radius $r$ around the jet axis. The integrated shapes are by definition normalized such that $\Psi(r/R = 1) = 1$. By definition $\Psi(0) = 0$. In this analysis, we do not consider particles outside the jet cone radius. In Monte Carlo simulation, $b$-jets are defined as jets which have at least one $b$-quark inside the jet cone.
%
\section{Event Selection}\label{sec:data}
In this section the online event selection criteria (triggers) used for this measurement are described. This is followed by a description of the method used to enhance the heavy-flavor fraction of jets by requiring there to be a reconstructed displaced vertex inside the jet cone. The selections (cuts) applied to the events are then introduced and finally the correction factors applied to the jet $p_T$ are described.\\
This paper presents results for central jets, $|y|\leq 0.7$, in a $p_T$ range from 52 to 300~GeV/$c$. Four different datasets are used. Events were collected that satisfy the conditions required by the inclusive jet trigger, with different  minimum transverse energy thresholds for the different datasets. Each dataset is defined by a unique trigger path that has unique requirements at each of the three stages of the trigger (Trigger Levels L1, L2 and L3) . The inclusive jet triggers select events based solely on the energy deposits in the calorimeters, with four different thresholds on the jet transverse energy, $E_T$ (see Table~\ref{tab:triggers}). Due to limited bandwidth, only a fraction of the events in each trigger path is accepted~\cite{prescales}. In the first-level trigger, a single trigger tower~\cite{trigger_tower} with $E_T$ above 5 or 10~GeV, depending on the trigger path, is required. In the second-level trigger, calorimeter tower clusters are formed around the selected trigger towers. Starting from the initial first-level tower, adjacent towers are associated to it if their $E_T$ is above 1~GeV, thus forming clusters of calorimeter towers. The events are required to have at least one second-level trigger cluster with $E_T$  above a given threshold, which varies between 15 and 90~GeV for the different trigger paths. In the third-level trigger, jets are reconstructed using the CDF Run I cone algorithm~\cite{cdf_1_algo}, and the events are required to have at least one jet with $E_T$ above selected thresholds between 20 and 100~GeV.\\
 The inclusive jet  triggers are not fully efficient at the trigger threshold with respect to offline reconstructed jets. To avoid any trigger bias, events in each dataset are only considered when the trigger efficiency is above $99\%$. These trigger efficiency thresholds are shown in the last column of Table~\ref{tab:triggers} where the jet transverse momentum is quoted after corrections for detector effects, as discussed later in this section. The lowest $p_T$ for this analysis, 52~GeV, is determined by the trigger threshold for the lowest $p_T$ inclusive jet trigger used.\\
 \begin{table}[htp]
\begin{small}
\begin{center}
\begin{tabular}{ccccc} \hline\hline
Trigger Path & L1 & L2 & L3  & 99\% trig. eff. \\
 & Tower & Cluster &Jet & Offline \\
  & $E_T$ [GeV] & $E_T$ [GeV] & $E_T$ [GeV]  & $p_T$ [GeV/$c$]\\
\hline
JET 20 & 5 & 15 & 20 & 52\\
JET 50 & 5 & 40 & 50 &80 \\
JET 70 & 10 & 60 & 70 &104\\
JET 100 & 10 & 90 & 100 &142\\
\hline
\hline
\end{tabular}
\caption{Summary of trigger paths and trigger thresholds used in each of the three CDF trigger levels. The last column shows the final offline cuts applied to the corrected jets after the 99\% trigger efficiency requirement has been applied.}\label{tab:triggers}
\end{center}
\end{small}
\end{table}
The inclusive jet datasets are dominated by light quark and gluon jets. The $b$-jet content is enhanced by using a secondary vertex tagger~\cite{secvtx}. This tagging algorithm exploits the long lifetime of the $b$-hadrons. Due to the large relativistic boost of the particles, most $b$-hadrons travel a few millimeters before decaying. The tagging algorithm is based on the reconstruction of a displaced, or secondary, vertex using the reconstructed charged particle trajectories, tracks, contained within a cone of 0.4, in $(y,\phi)$-space, around the jet axis. Despite the jet cone size being 0.7, the cone for finding displaced tracks, the tagging cone, is kept at 0.4 because the direction of the heavy-flavor hadron and its decay products tend to be close to the direction of the jet axis. Using a larger cone size for the tracks would lead to an increase in misidentified secondary vertices. The tagging algorithm implementation is relatively complex and is described in~\cite{secvtx}; a brief overview is given here. Before attempting to reconstruct a displaced vertex, several quality cuts are applied to the tracks inside the tagging cone in order to reduce the number of tracks that are not well measured. Tracks are then ordered according to quality criteria, including their distance of closest approach to the primary vertex, $d_0$. Starting from the track with the highest quality, an attempt at creating a displaced vertex is made with the next best track. The attempt succeeds or fails depending on the quality of the fit. If the attempt succeeds, all other tracks whose $d_0$ significance~\cite{foot:d0_sig} with respect to the displaced vertex is below a predefined limit are attached to it. If the attempt fails, the highest quality track is associated to the third best track and the process starts over. If no vertex is found with the highest quality track, the second highest quality track is considered and attempts are made to reconstruct a displaced vertex with the next best track. This loop continues until either a good displaced vertex is found or there are no more pairs of tracks that pass the selection criteria. If a good displaced vertex is found, a final cut is applied on the significance of the two-dimensional projection along the jet axis of the distance between the primary and secondary vertex locations~\cite{foot:lxy}. The main limitation of this tagging method is the fact that $c$-hadrons can have decay lengths similar to those of $b$-hadrons. Another limitation is the finite resolution of both the primary and secondary vertex location.\\
The following list summarizes all the cuts applied to the events in the samples:\\
One, and only one, primary vertex with z-component, $z_{\mathrm{vtx}}$, in the region $|z_{\mathrm{vtx}}| \leq50$~cm to ensure good secondary vertex reconstruction and to reject events with multiple $p\bar{p}$ interactions; 
missing transverse energy significance of the event, $\not \! \! E_T/ \sqrt{E_T}$~\cite{foot:met},  must be smaller than a given threshold, varying from 3.5  to 7.0~GeV$^{1/2}$ depending on the trigger used, in order to reject cosmic rays that enter the detector as well as to reject beam induced backgrounds;\\
$|y_\mathrm{jet}|\leq 0.7$ where the secondary vertex tagging algorithm is best understood;\\
for the $b$-jet enriched samples only, jets are required to have a secondary vertex tag~\cite{secvtx}.\\
An average jet correction is applied to correct the jet $p_T$ to the hadron level, i.e. to remove any bias due to the detector. It is calculated by matching, in the Monte Carlo simulation, hadron and calorimeter level jets in ($y-\phi$)-space. The jet $p_T$ is obtained by running the jet reconstruction algorithm over the final state hadrons and energy deposits in the calorimeter towers, respectively. The hadron level jet $p_T$ is plotted as a function of the calorimeter jet $p_T$. The scatter plot obtained is fitted, using a chi-squared minimization, to a fourth order polynomial which is then used to correct the transverse momentum of each measured jet in data. The increase in $p_T$ due to this correction is on the order of $20\%$ at low $p_T$ and $10\%$ at higher $p_T$. The binning used for this analysis and for all plots shown is in corrected jet $p_T$, referred to in the remainder of this paper simply as the jet $p_T$. This is an absolute correction to the jet that does not affect the shape of a given jet in the sample, only its total $p_T$ and hence which $p_T$ bin it belongs to. The corrections to the jet shape variables are discussed in Sec.~\ref{sec:had_corr}.

\section{Monte Carlo Models}\label{sec:mc}
This section introduces the two leading order MC models to which the measured $b$-jet shapes are compared: {\sc pythia} Tune A and {\sc herwig}. In both MC models, the parton shower is implemented to leading log. For both MC models, inclusive dijet samples were produced (msel = 1) using the CTEQ5L parton distribution functions~\cite{cteq5l}.\\
 {\sc pythia} Tune A is based on {\sc pythia} v6.203~\cite{pythia} which has been tuned to the CDF Run I underlying event. It is worth noting that the underlying event tuning is found to be important for a good description of the inclusive jet shape~\cite{shapes_incl}. This underlying event tune is widely used in CDF analyses~\cite{shapes_incl}. The final-state parton shower model of {\sc pythia} used for this analysis is carried out in the {\sc jetset } part of the code~\cite{jetset}; this time-like evolution is computed using splitting functions and an ordering in virtuality scale, $Q^2$. The splitting functions, $P_{a\rightarrow bc }(z)$, describe the probability that a parton $a$ splits into two partons $b$ and $c$ where parton $b$ carries away a momentum fraction $z$ of the initial parton. The virtuality scale in {\sc pythia} is equal to the four-momentum squared of the branching parton. The ordering in virtuality scale means that for initial state radiation, $Q^2$ increases as the process unfolds from the incoming primary parton as it approaches the hard collision. For final state radiation, $Q^2$ decreases going from the energetic outgoing parton to the final shower. Decreasing $Q^2$ in the final state is similar to the decreasing angular ordering used by {\sc herwig}. The parton shower is terminated at $Q_0$ of 1~GeV/$c$ for QCD branchings. For the non-perturbative fragmentation and hadronization processes, our version of {\sc pythia} uses the default Lund string model, as implemented in {\sc jetset}. In this model, the long-range confinement forces are allowed to distribute the energies and flavors of a parton configuration among a collection of primary hadrons which subsequently may decay further. \\
The modeling of the decay of $b$- and $c$-hadrons is based on the program {\sc qq}~\cite{qq}, the decay tables of which are periodically updated with the latest experimental information.  When {\sc pythia} produces a $b$-hadron in a jet, {\sc qq} is called to handle the decay. For some decay modes the exclusive final states are only obtained after fragmentation of the quarks, which requires a return to {\sc pythia}.\\
 The second model to which the experimental results are compared is {\sc herwig} v6.505~\cite{herwig}. In the perturbative parton showering process, the main difference with respect to {\sc pythia} is the use of angular ordering of successive emissions, which simulates the color flow of the subprocesses. For final state radiation, the angle of the radiated gluon relative to the parent parton direction decreases as the process unfolds. The maximum angle is determined by the elementary subprocess and is due to interference among gluons. {\sc herwig} uses the cluster model for the fragmentation into hadrons~\cite{cluster}. This model, which is independent of the initial hard process and of the energy, is intended to disrupt as little as possible the event structure established by the parton showering. Color-neutral clusters are formed that decay into the observed hadrons. The decay of heavy flavored hadrons is treated in {\sc herwig} in the same way as for all other unstable particles, according to the exponential decay law using the mean lifetime of the particle. This version of {\sc herwig} does not include multiple parton interactions, which is thought to be the main reason why the inclusive jet shapes agree better with {\sc pythia} Tune A than {\sc herwig}~\cite{shapes_incl}. 
%

\section{B-jet Shape Unfolding}\label{sec:unfolding}
In this section, the analysis methodology is presented which makes it possible to extract the shapes of $b$-jets from samples of inclusive and tagged jets. First the unfolding equations are described, then in Sec.~\ref{sec:f1b_fraction} the effect on these equations of the fraction of $b$-jets that contain more than one $b$-quark is discussed. The next sub-sections present how the different components of the unfolding equations are measured; Section~\ref{sec:raw} presents the measurement of the inclusive and tagged samples; Section~\ref{sec:purity} presents the method used to extract the $b$-jet fraction in the tagged samples; the method used to account for the bias to the jet shapes due to tagging is presented in Sec.~\ref{sec:biases}; finally the corrections applied to the measured $b$-jet shapes in order to obtain a detector independent result are presented in Sec.~\ref{sec:had_corr}.\\
The jet shape variables, defined in equation~\ref{equ:int_shape_def}, are computed in data using the calorimeter towers with $p_{T} > 0.1$~GeV/$c$. The use of calorimeter towers with $p_T > 0.5$~GeV/$c$ as well as the use of tracks with $p_T>0.5$~GeV/$c$ are considered in the study of the systematic uncertainties. The sample of tagged jets used for this analysis does not contain only $b$-jets but also background jets, called non$b$-jets, that do not contain any $b$-quarks. The non$b$-jets contain $c$-jets,  light flavor-jets and gluon-jets, where the gluon does not fragment into a $b\bar{b}$-pair. The purity, $p_\mathrm{b}$, is defined as the fraction of tagged jets that are $b$-jets. The detector level jet shapes for tagged jets, $\Psi^\mathrm{tag}_\mathrm{det} $, will thus be given by
\ifprd
\begin{eqnarray}
\Psi^\mathrm{tag}_\mathrm{det}(r/R) & = & p_\mathrm{b} \Psi^\mathrm{tag \, b}_\mathrm{det}(r/R) \nonumber \\
 & &  + (1-p_\mathrm{b})\Psi^\mathrm{tag\, nonb}_\mathrm{det}(r/R), \label{equ:det_shapes_1}
\end{eqnarray}
\else
 \begin{linenomath*}
\begin{equation}
\Psi^\mathrm{tag}_\mathrm{det}(r/R) =p_\mathrm{b} \Psi^\mathrm{tag \, b}_\mathrm{det}(r/R) + (1-p_\mathrm{b})\Psi^\mathrm{tag \, nonb}_\mathrm{det}(r/R), \label{equ:det_shapes_1}
\end{equation}
\end{linenomath*}
\fi
 where $\Psi^\mathrm{tag \, b}_\mathrm{det}$ and $\Psi^\mathrm{tag\,nonb}_\mathrm{det}$ are the $b$- and non$b$-jet contributions to the measured tagged jet shape.\\
  The use of the secondary vertex tagging method to enhance the heavy flavor content of the sample biases the measured jet shapes due to the fact that the secondary vertex reconstruction algorithm requires jets with clean and well defined tracks. A bias term, dependent on the distance from the jet axis, $r$, and on the $p_T$ of the jet must thus be added to correct for this effect without affecting the overall normalization of the integrated jet shapes. This bias is different for $b$-jets and non$b$-jets. In the case of non$b$-jets, the bias terms also take into account the enhanced fraction of $c$-jets in the tagged jet sample. The bias terms, $b_\mathrm{b}(r/R)$ and $b_\mathrm{nonb}(r/R)$ for $b$- and non$b$-jets respectively, are defined in MC simulation such that
  \ifprd
\begin{eqnarray}
\Psi^\mathrm{tag \,b}_\mathrm{det}(r/R) & = & b_\mathrm{b}(r/R) \Psi_\mathrm{det}^\mathrm{b}(r/R) \,\,\mathrm{and}\\
\Psi^\mathrm{tag \, nonb}_\mathrm{det}(r/R)& =&  b_\mathrm{nonb}(r/R) \Psi_\mathrm{det}^\mathrm{nonb}(r/R)\label{equ:bias},
\end{eqnarray}
\else
\begin{linenomath*}
\begin{equation}
\Psi^\mathrm{tag \, b}_\mathrm{det}(r/R) = b_\mathrm{b}(r/R) \Psi_\mathrm{det}^\mathrm{b}(r/R) \,\, , \,\,\Psi^\mathrm{tag \, nonb}_\mathrm{det}(r/R) = b_\mathrm{nonb}(r/R) \Psi_\mathrm{det}^\mathrm{nonb}(r/R)\label{equ:bias},
\end{equation}
\end{linenomath*}
\fi
where the $\Psi_\mathrm{det}(r/R)$ terms represent the detector level shapes, as obtained from MC simulation, before any tagging requirements and $\Psi^\mathrm{tag}(r/R)$ are those obtained from MC simulation after the tagging requirement is applied. The bias terms are computed separately for each $p_T$ bin.\\
Combining Eq.~(\ref{equ:det_shapes_1}) with the definition of the bias terms from Eq.~(\ref{equ:bias}), the measured detector level jet shapes for tagged jets can be re-written as
\ifprd
\begin{eqnarray}
\Psi^\mathrm{tag}_\mathrm{det}(r/R) &= & p_\mathrm{b}  b_\mathrm{b}(r/R) \Psi_\mathrm{det}^\mathrm{b}(r/R) 
\nonumber \\
& & + (1-p_\mathrm{b}) b_\mathrm{nonb}(r/R) \Psi_\mathrm{det}^\mathrm{nonb}(r/R), \label{equ:det_shapes_2}
\end{eqnarray}
\else
 \begin{linenomath*}
\begin{equation}
\Psi^\mathrm{tag}_\mathrm{det}(r/R) =p_\mathrm{b}  b_\mathrm{b}(r/R) \Psi_\mathrm{det}^\mathrm{b}(r/R) + (1-p_\mathrm{b}) b_\mathrm{nonb}(r/R) \Psi_\mathrm{det}^\mathrm{nonb}(r/R), \label{equ:det_shapes_2}
\end{equation}
\end{linenomath*}
\fi
Rearranging Eq.~(\ref{equ:det_shapes_2}), the detector level measurement of the $b$-jet shape is defined as
\ifprd
\else
\begin{linenomath*}
\fi
\begin{equation}
\Psi_\mathrm{det}^\mathrm{b}(r/R) = \frac{ \Psi^{tag}_\mathrm{det}(r/R)- (1 - p_\mathrm{b}) b_\mathrm{nonb}(r/R) \Psi_\mathrm{det}^\mathrm{nonb}(r/R)} {p_\mathrm{b}  b_\mathrm{b}(r/R) }. \label{equ:b_shape_det_3}
\end{equation}
\ifprd
\else
\end{linenomath*}
\fi
It is also necessary to correct the $b$-jet shapes for detector effects, i.e. to remove all influence of the tracker or calorimeters on the measurement. The corresponding hadron level correction factors, $C^\mathrm{had}(r/R)$, are defined as
\ifprd
\begin{eqnarray}
\Psi_\mathrm{had}^\mathrm{b}(r/R) &=& C^\mathrm{had}(r/R)  \Psi_\mathrm{det}^\mathrm{b}(r/R) \label{equ:had_corr},
\end{eqnarray}
\else
\begin{linenomath*}
\begin{equation}
\Psi_\mathrm{had}^\mathrm{b}(r/R) = C^\mathrm{had}(r/R) \Psi_\mathrm{det}^\mathrm{b}(r/R) \label{equ:had_corr},
\end{equation}
\end{linenomath*}
\fi
where in MC simulation the shapes are computed both at the detector level, $ \Psi_\mathrm{det}^\mathrm{b}(r/R)$, and using the final state particles, $\Psi_\mathrm{had}^\mathrm{b}(r/R)$. These definitions of the tagging bias and hadron level correction factors ensure the proper normalization of the jet shapes at each stage of the unfolding procedure. They are defined such that $C^\mathrm{had}(r/R=1) =1$, $ b_\mathrm{b}(r/R=1)=1$ and  $b_\mathrm{nonb}(r/R=1) =1$.\\
The final equation used to obtain the hadron level $b$-jet shape is obtained by combining Eq.~(\ref{equ:b_shape_det_3}) and Eq.~(\ref{equ:had_corr}). It can be written as
\ifprd
\begin{widetext}
\begin{eqnarray}
\Psi_\mathrm{had}^\mathrm{b}(r/R) &=& C^\mathrm{had}(r/R)  \frac{ \Psi^{tag}_\mathrm{det}(r/R)- (1 - p_\mathrm{b}) b_\mathrm{nonb}(r/R) \Psi_\mathrm{det}^\mathrm{nonb}(r/R)} {p_\mathrm{b}  b_\mathrm{b}(r/R) }, \label{equ:b_shape_det_4}
\end{eqnarray}
\end{widetext}
\else
\begin{linenomath*}
\begin{equation}
\Psi_\mathrm{had}^\mathrm{b}(r/R) = C^\mathrm{had}(r/R) \frac{ \Psi^{tag}_\mathrm{det}(r/R)- (1 - p_\mathrm{b}) b_\mathrm{nonb}(r/R) \Psi_\mathrm{det}^\mathrm{nonb}(r/R)} {p_\mathrm{b}  b_\mathrm{b}(r/R) }, \label{equ:b_shape_det_4}
\end{equation}
\end{linenomath*}
\fi
where $ \Psi^\mathrm{tag}_\mathrm{det}(r/R)$ is the measured jet shape for the tagged jet sample and $\Psi_\mathrm{det}^\mathrm{nonb}(r/R)$ is the measured inclusive jet shape, described in Sec.~\ref{sec:raw}. The other parameters of this equation are discussed in the following sub-sections.

\subsection{Single $b$-quark jet content}\label{sec:f1b_fraction}
\begin{figure*}[h]
  \begin{center}
           \includegraphics[width=12.0cm,clip=]{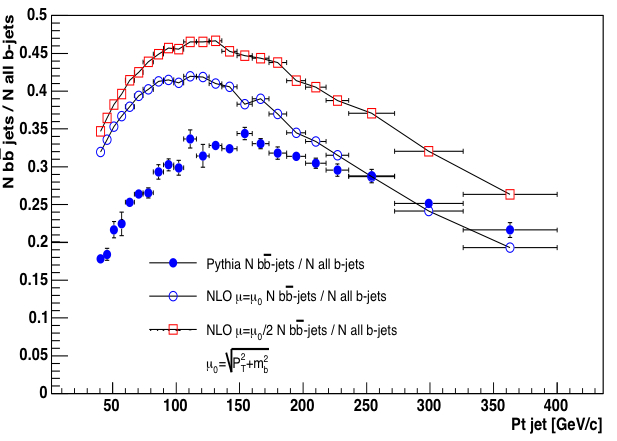}
      \caption{Fraction of $b$-jets that contain more than one $b$-quark inside the same jet cone. The {\sc pythia} Tune A MC simulation predictions are compared to NLO calculations for two different hadronisation and factorisation scales. The NLO calculations are shown binned with the same binning as used for the {\sc pythia} Tune A MC.
      \label{fig:f1b_comp}}
  \end{center}
\end{figure*}

Many of the distributions used for the extraction of the b-jet shapes described in the previous section are expected to be different depending on whether the jets contain one or two $b$-quarks. This section describes how this is dealt with in this measurement. The parameters used in the unfolding are sensitive to the fraction of $b$-jets that contain a single $b$-quark, $f_\mathrm{1b}$. In leading order (LO) Monte Carlo, gluon splitting to $b \bar{b}$ pairs occurs as part of the fragmentation process and not in the matrix element. For most jets where gluons split to a $b\bar{b}$ pair, both $b$-quarks end up inside the same jet cone~\cite{heavy_flavor_summary}. A comparison between the fraction of $b$-jets with more than one $b$-quark inside the same jet cone, $1-f_\mathrm{1b}$, predicted by {\sc pythia} Tune A and the next to leading order (NLO) calculation  is shown in Fig.~\ref{fig:f1b_comp}~\cite{nlo_f1b}~\cite{nlo_f1b_2}. The maximum deviation between the {\sc pythia} Tune A and the NLO prediction with the factorization and renormalisation scales, $\mu$, equal to $\frac{\mu_0}{2}$ is on the order of 0.2. This particular choice of $\mu$ is motivated by the measurement of the inclusive b-jet cross section~\cite{nlo_f1b_2}. Before calculating any of the unfolding factors, the Monte Carlo samples are re-weighted to decrease the $f_\mathrm{1b}$ fraction by 0.2, in order to account for this underestimation of the gluon splitting fraction. This value was chosen for all $p_T$ bins as it corresponds to roughly the expected shift. For each of the $b$-jet shapes obtained from MC simulation: the unbiased $b$-jet shapes, the tagged $b$-jet shapes and the hadron level $b$-jet shapes, the shapes are extracted separately for single, $\Psi^\mathrm{1b}$, and double $b$-quark jets, $\Psi^\mathrm{2b}$. They are then combined using the new fraction of single $b$-quark jets as follows
\ifprd
\begin{eqnarray}
 \Psi^\mathrm{b} = (f_\mathrm{1b}-0.2) \Psi^\mathrm{1b}+ (1 - (f_\mathrm{1b}-0.2)) \Psi^\mathrm{2b}
\label{equ:f1b_new_comb}.
\end{eqnarray}
\else
\begin{linenomath*}
\begin{equation}
 \Psi^\mathrm{b} = (f_\mathrm{1b}-0.2) \Psi^\mathrm{1b}+ (1 - (f_\mathrm{1b}-0.2)) \Psi^\mathrm{2b}
\label{equ:f1b_new_comb}.
\end{equation}
\end{linenomath*}
\fi
This new lower $f_\mathrm{1b}$ fraction is used for the corrections to data in the remainder of the analysis, to obtain the secondary vertex mass templates, as well as the tagging biases and hadron level corrections to the $b$-jet shapes. We evaluate the systematic effect of this particular choice of correction to the $f_\mathrm{1b}$ fraction in Sec.~\ref{sec:sys}.

\subsection{Detector Level Jet Shapes}\label{sec:raw}
\begin{figure*}[h]
  \begin{center}
           \includegraphics[width=18.0cm,clip=]{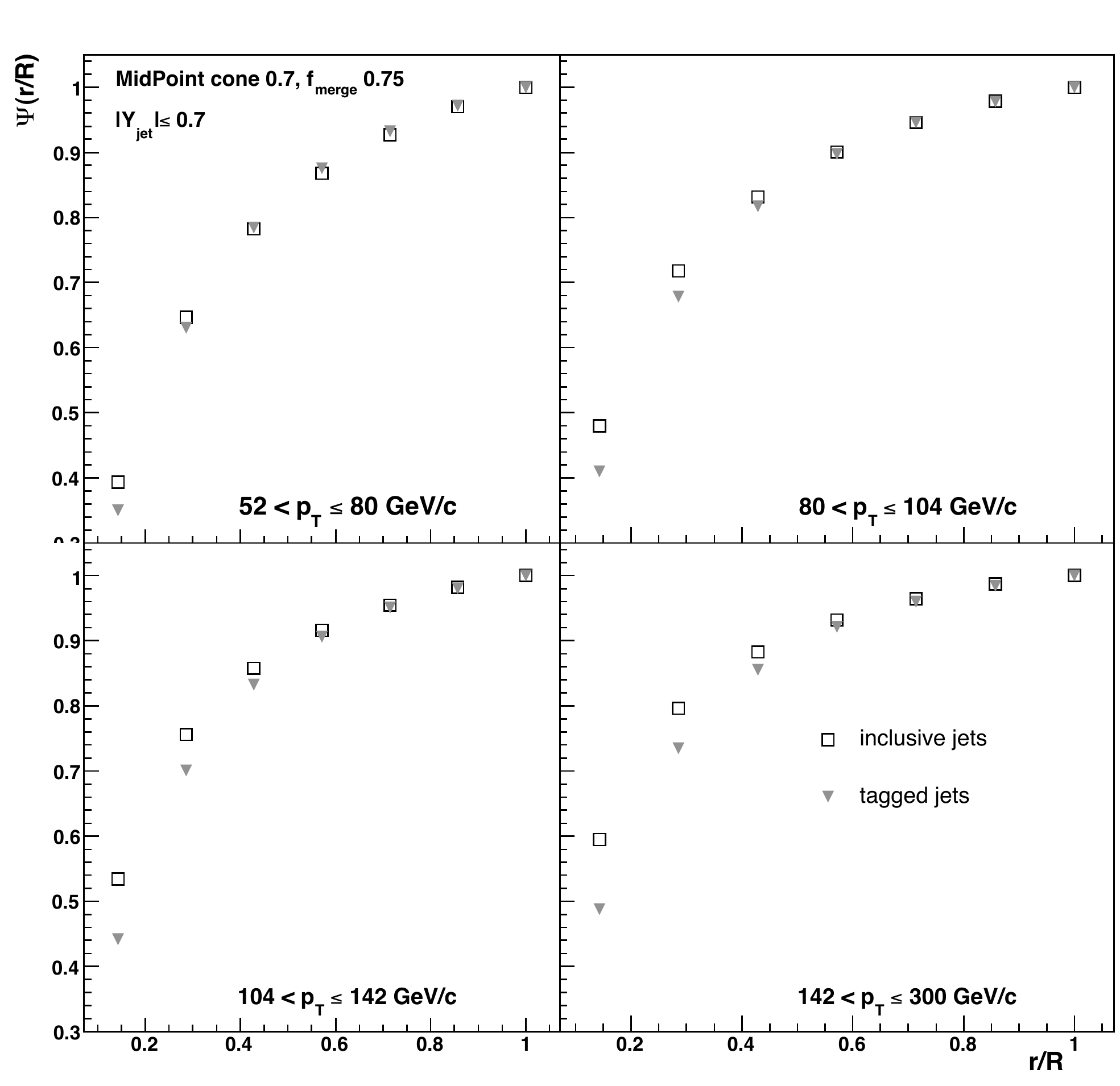}
      \caption{Measured detector level jet shapes in data for the tagged (grey full triangles) and inclusive (open squares) data samples for each $p_T$ bins. The jet algorithm used, MidPoint cone 0.7 $f_{merge}$ 0.75,  is described in Sec.~\ref{sec:midpoint}.
 \label{fig:shapes_incl_tag_data}     }
  \end{center}
\end{figure*}

In the jet samples considered for this analysis, the average multiplicity of calorimeter towers inside the jet increases slightly as the jet $p_T$ increases from 16 to 23 towers and is on average a little higher for tagged jets than for inclusive jets, which range from 13 to 18 towers. The multiplicity of tracks that pass all the selection cuts exhibits the same behavior. The variation is from 16 to 23 tracks for tagged jets, and 13 to 17 tracks for inclusive jets. Jets with a higher $p_T$ have more tracks due to the fact that the final number of particles is determined by the number of times the initial parton radiated another parton, before it hadronizes. Jets with higher $p_T$ have on average a higher initial $Q^2$ which means a larger range over which to radiate more particles.\\
The integrated jet shape is computed for each jet $i$ in the sample as
\ifprd
\else
\begin{linenomath*}
\fi
\begin{equation}
\Psi_i(r/R) = \frac{\sum^\mathrm{tow}_{j}{p_T^{j}(x \leq r/R)}}{\sum^\mathrm{tow}_{k}{p_T^{k}(x \leq 1)}}\label{equ:jet_shape_calc},
\end{equation}
\ifprd
\else
\end{linenomath*}
\fi
where the sum is over all towers that satisfy the conditions of belonging to the jet and that the fractional distance to the jet axis, $x$, must be less than $r/R$ in the numerator and 1 in the denominator. The use of this denominator ensures the correct shape normalisation for each jet. For each $p_T$ bin, the average jet shape is computed from the jet shapes of all jets in that sample
\ifprd
\else
\begin{linenomath*}
\fi
\begin{equation}
\Psi(r/R) = <\Psi_i(r/R)>\label{equ:jet_shape_average}
\end{equation}
\ifprd
\else
\end{linenomath*}
\fi
The tagged jet shapes, $\Psi^\mathrm{tag}(r/R)$, are defined as the average jet shapes, measured at the detector level, of all tagged jets in the samples.\\
Given the very low fraction of $b$-jets in inclusive jet production, estimated from the MC to be less than $4\%$, it is possible to approximate the non$b$-jet shapes to those of the inclusive jet shapes, before any tagging requirements. The assumption is that $\Psi_\mathrm{det}^\mathrm{nonb}(r/R) \approx \Psi_\mathrm{det}^\mathrm{incl}(r/R)$. The difference between these shapes, in {\sc pythia} Tune A Monte Carlo, is negligible with a maximum difference of less than $0.5\%$. No systematic uncertainty is therefore related to the use of this approximation.\\
Both the inclusive jet shapes, $\Psi_\mathrm{det}^\mathrm{incl}(r/R)$ and the tagged jet shapes,  $\Psi_\mathrm{det}^\mathrm{tag}(r/R)$, are measured from data in each of the four $p_T$ bins. These detector level jet shapes are shown in Fig.~\ref{fig:shapes_incl_tag_data} for all four $p_T$ bins. Comparing these inclusive and tagged jet shapes shows that there are significant differences in the measured shapes, yielding confidence that this measurement is possible. But at this stage, it is not possible to conclude anything about the shapes of $b$-jets as the displaced vertex requirement biases the observed tagged shapes, as described in Sec.~\ref{sec:biases}.

\subsection{Purity}\label{sec:purity}
\begin{figure*}[htp]
  \begin{center}
 \ifprd
    \includegraphics[width=11.0cm,clip=]{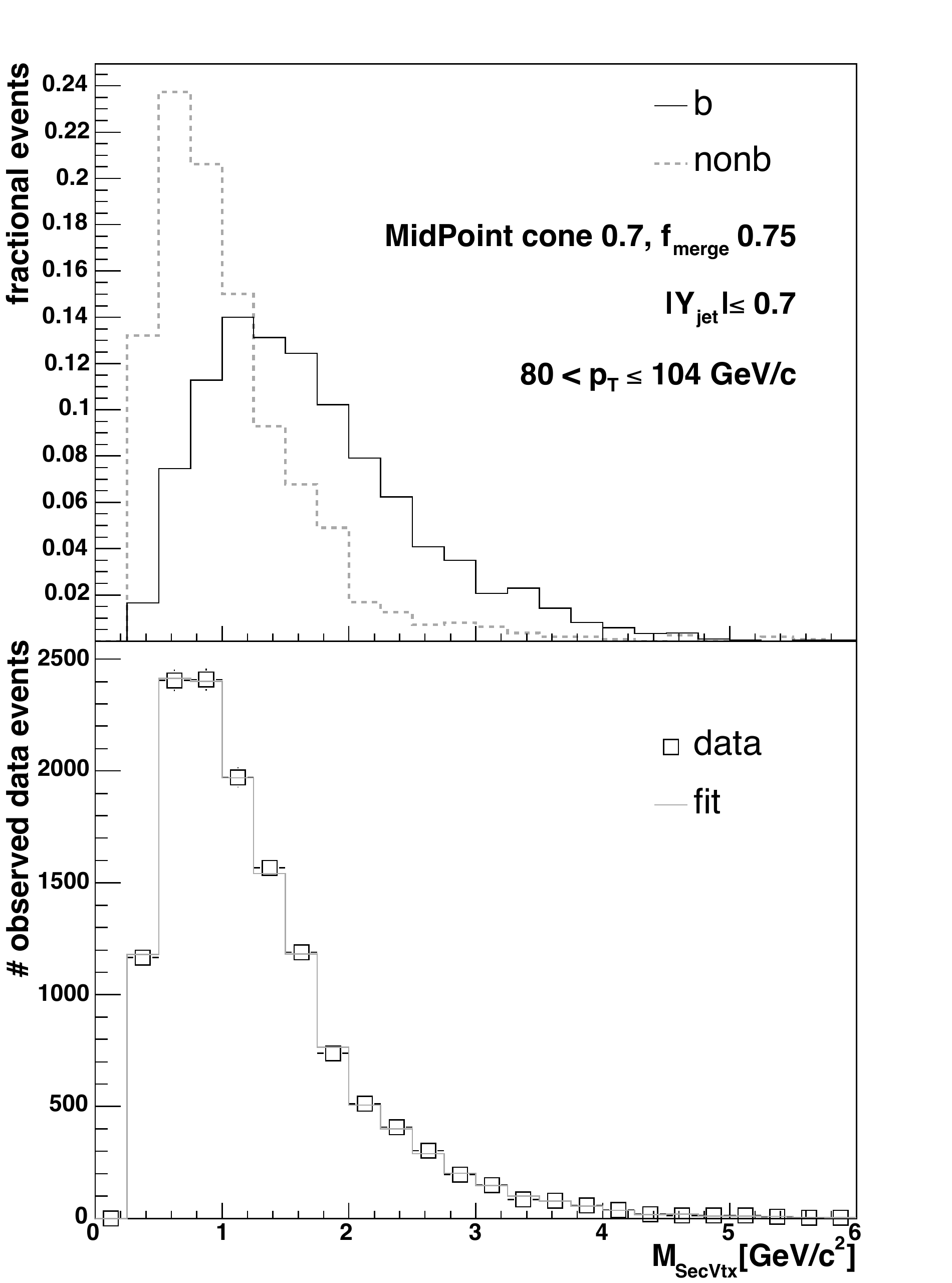}
\else
    \includegraphics[width=9.0cm,clip=]{purity_incl_py_1.pdf}
 \fi
   \includegraphics[width=11.0cm,clip=]{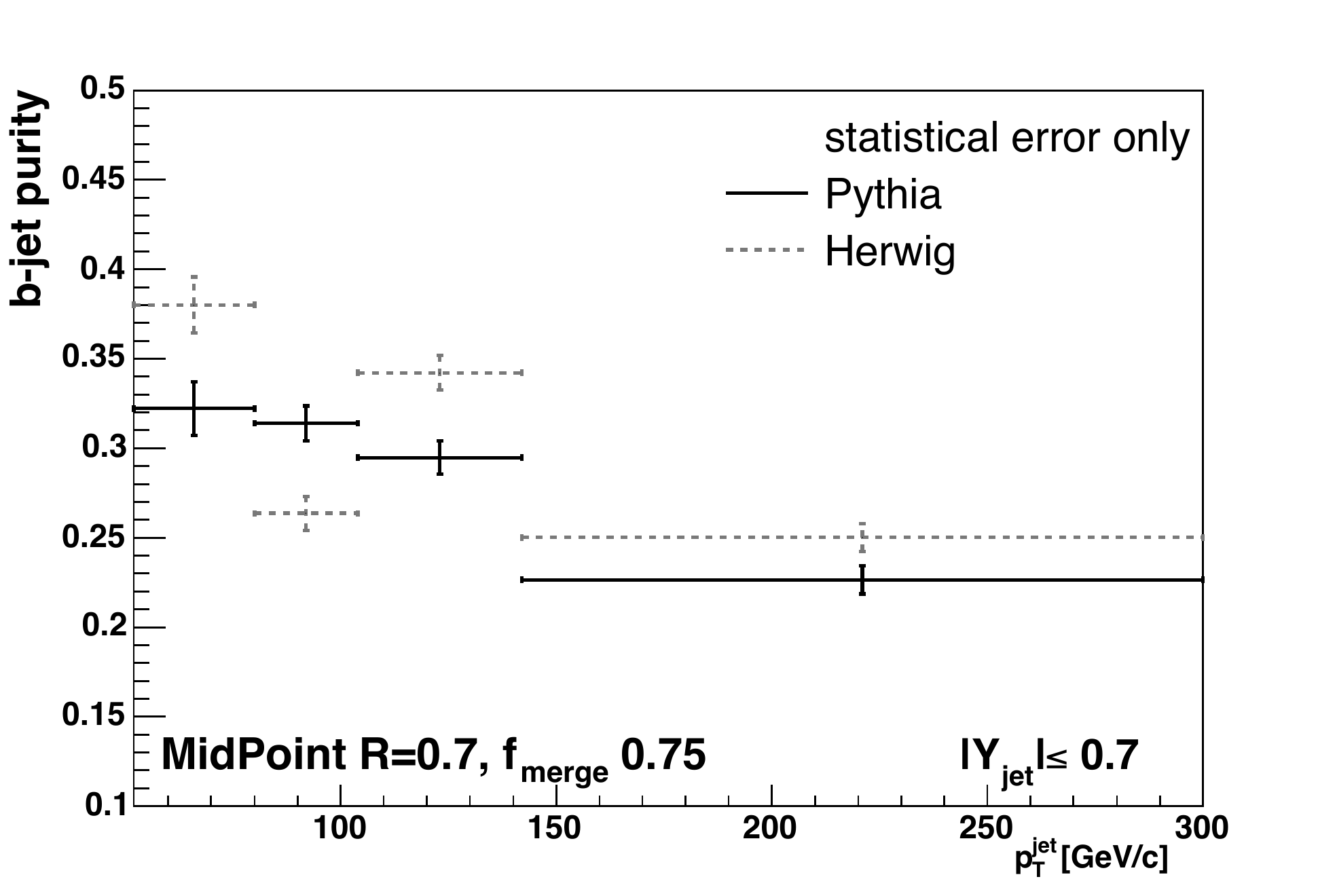}
    \caption{(top) Normalized secondary vertex mass distributions for $b$- (black full line) and non$b$-jets (grey dashed line) as obtained from {\sc pythia} Tune A. (middle) Secondary vertex mass distribution in data (black open squares) compared to the fitted distribution (histogram) for the second $p_T$ bin. (bottom) Extracted $b$-jet purity in data as a function of jet $p_T$ using the templates obtained from {\sc pythia} Tune A (black full lines) and from {\sc herwig} (grey dashed lines). The error bars indicate the statistical uncertainties only.\label{fig:secvtx_fit}}
  \end{center}
\end{figure*}

The fractions of $b$-jets in the tagged jet samples are extracted from fits to invariant mass distributions for $b$- and non$b$-jets, calculated in {\sc pythia}, from the charged tracks forming the secondary  vertex. It is not possible to reconstruct the full hadron invariant mass mainly because of the presence of neutral particles in the $b$-hadron decays that are not detected in the tracking detectors. Nevertheless, the distribution of the invariant mass of the tracks associated with the secondary vertex, referred to as the secondary vertex mass, is significantly different for real heavy flavored jets than for mis-reconstructed light flavored or gluon jets. Using the Monte Carlo samples, distributions of the secondary vertex masses for tagged jets are obtained for each $p_T$ bin, separately for $b$- and non$b$-jets. The non$b$-jet distribution is a combination of real displaced vertices from $c$-jets as well as light flavor jets where a secondary vertex was mistakenly identified by the secondary vertex algorithm. As an example, the secondary vertex mass distributions, as obtained from {\sc pythia} Tune A, for the second $p_T$ bin, from 80 to 104~GeV/$c$, are shown in the top two plots of Fig.~\ref{fig:secvtx_fit}. The measured distribution in the data is fitted to the $b$- and non$b$-templates, using a binned $\chi^2$ minimization method, to find the most probable fraction of jets that are $b$-jets, as shown in the second plot of Fig.~\ref{fig:secvtx_fit} for the second $p_T$ bin. The fit describes the data very well in all $p_T$ bins. Figure~\ref{fig:secvtx_fit} (bottom) shows the extracted purity, $p_\mathrm{b}$, as a function of the $p_T$ of the jets as obtained by fitting the data with templates from {\sc pythia} Tune A on the one hand and from {\sc herwig} on the other. 

\subsection{Biases Due to Secondary Vertex Tagging}\label{sec:biases}
The requirement that the jets be tagged by the secondary vertex algorithm introduces a bias in the measured jet shapes. These biases are different for each $p_T$ bin, each bin in $r$ and different for $b$- and non$b$-jets. The bias terms are defined as the ratios, as obtained from the Monte Carlo samples, between the tagged and the unbiased jet shapes for $b$- and non$b$-jets separately
\ifprd
\else
\begin{linenomath*}
\fi
\begin{equation}
b_\mathrm{b}(r/R) = \frac{\Psi^\mathrm{tag \,b}_\mathrm{MC}(r/R)}{\Psi_\mathrm{MC}^\mathrm{b}(r/R)} \, ,
\end{equation}
\ifprd
\else
\end{linenomath*}
\fi
\ifprd
\else
\begin{linenomath*}
\fi
\begin{equation}
 b_\mathrm{nonb}(r/R) = \frac{\Psi_\mathrm{MC}^\mathrm{nonb \, tag}(r/R)}{\Psi_\mathrm{ MC}^\mathrm{nonb \, incl}(r/R)}.\label{equ:bias_2}
\end{equation}
\ifprd
\else
\end{linenomath*}
\fi
The bias term for non$b$-jets takes into account the increased fraction of $c$-jets in the tagged jet sample. The maximum bias for $b$-jets is on the order of $8\%$ and for non$b$-jets is on the order of $18\%$.\\
In this case the $b$-tagging efficiency is not relevant as we are interested in the distortions to the average jet shape arising from the tagging and not in absolute cross sections.

\subsection{Hadron Level Corrections}\label{sec:had_corr}
The hadron level correction factors that remove the influence of the detector on the measured $b$-jet shapes, $C^\mathrm{had}(r/R)$, are evaluated from the Monte Carlo samples for each bin in $r$ and each bin in $p_T$ and are defined as
\ifprd
\else
\begin{linenomath*}
\fi
\begin{equation}
C^\mathrm{had}(r/R) = \frac{\Psi_\mathrm{had \, MC}^\mathrm{b}(r/R)}{\Psi_\mathrm{det \, MC}^\mathrm{b}(r/R)},\label{equ:had_corr_2}
\end{equation}
\ifprd
\else
\end{linenomath*}
\fi
where $\Psi_\mathrm{det \, MC}^\mathrm{b}(r/R)$ are the Monte Carlo simulated $b$-jet shapes computed at the detector level, and $\Psi_\mathrm{had \, MC}^\mathrm{b}(r/R)$ are the Monte Carlo simulated $b$-jet shapes computed using final state hadrons. These correction factors are on the order of $3\%$ at most.

\section{Systematic Uncertainties}\label{sec:sys}
\begin{figure*}[h]
  \begin{center}
\ifprd
    \includegraphics[width=18.0cm,clip=]{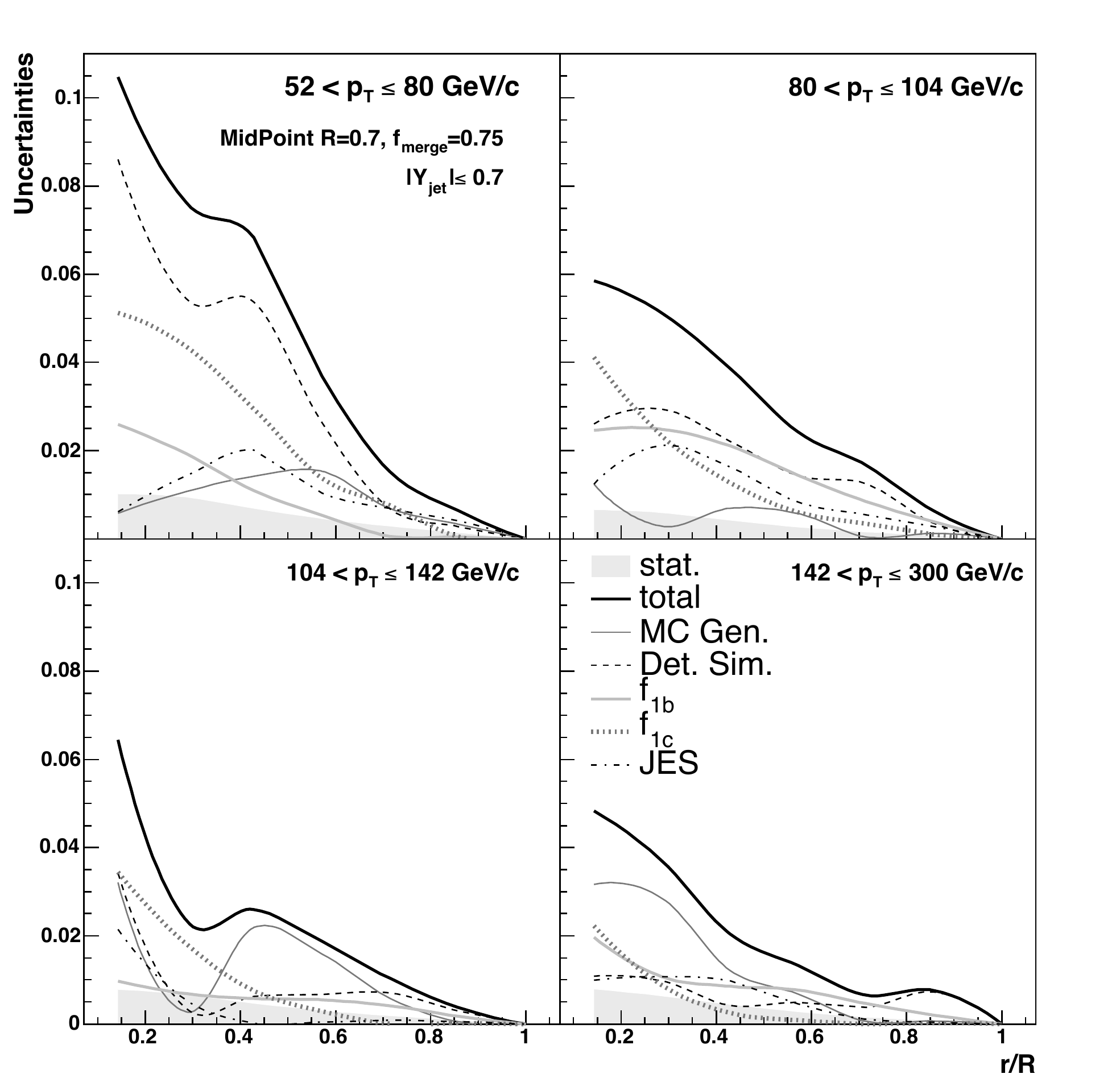}
    \else
    \includegraphics[width=15.0cm,clip=]{final_systematics.pdf}
\fi
    \caption{Total uncertainties on the integrated $b$-jet shape measurements for each of the four jet $p_T$ and $r$ bin (thick black lines). Also shown are the statistical uncertainties (grey bands) and the five dominant sources of systematic uncertainties: dependence on the particular MC model for the unfolding (thin black line), dependence on the detector simulation description (dashed thin black line), dependence on the single $b$-quark, $f_\mathrm{1b}$,  (thick grey line) and single $c$-quark, $f_\mathrm{1c}$ (dotted thick grey line) jet fractions in MC simulation, and the dependence on the jet energy scale, JES (dot-dashed black line) . The uncertainties are added in quadrature. \label{fig:total_sys}}
  \end{center}
  \end{figure*}

The different sources of systematic uncertainties for this measurement are described in this section.\\
To account for the sensitivity of the unfolding method to the variation of the $f_\mathrm{1b}$ fraction, the fraction is decreased by 0.5. The difference in the measured $b$-jet shapes when using the $f_\mathrm{1b} - 0.5$ instead of the default $f_\mathrm{1b} - 0.2$, discussed in Sec.~\ref{sec:f1b_fraction}, is taken as a systematic uncertainty.\\
The samples of tagged jets that are from non$b$-jets contain a significant fraction of $c$-jets; this comes from the fact that $c$-jets have real displaced vertices that are similar to those of $b$-jets and often get tagged by the secondary vertex algorithm. This is unlike the light jets where the reconstructed secondary vertex is not a true displaced vertex. The fraction of non$b$-jets in the tagged jet sample that are $c$-jets is taken from the Monte Carlo predictions and is found to be between $25\%$ and $50\%$, depending on the $p_T$ of the sample. From secondary vertex mass fits separating $b$-, $c$- and light-jets into three independent templates, a systematic uncertainty of $5\%$ is assigned to the $c$-jet content of the non$b$-jets. This uncertainty affects both the secondary vertex mass fit and the tagging bias for non$b$-jets.\\
The $c$-jets in the non$b$-jet sample are expected to have a somewhat similar behaviour to the $b$-jets, in particular regarding gluon splitting to $c\bar{c}$ pairs. Similarly to the $f_\mathrm{1b}$ fraction for $b$-jets, one can define the fraction of $c$-jets containing only one $c$-quark, $f_\mathrm{1c}$. This fraction might also be overestimated in the LO Monte Carlo as one would expect $f_\mathrm{1c}$ to increase if $f_\mathrm{1b}$ were to increase. In order to investigate the effect of a possible underestimation of this effect, the $f_\mathrm{1c}$ fraction is decreased by $0.2$. This change affects both the tagging bias for non$b$-jets as well as the templates used in the secondary vertex mass fit. The difference in the measured $b$-jet shapes when varying the $f_\mathrm{1c}$ fraction is taken as a systematic uncertainty.\\
In order to evaluate the effect of using a particular set of Monte Carlo models for the fragmentation, hadronisation and underlying event, the whole unfolding is performed using {\sc herwig} Monte Carlo samples instead of the {\sc pythia} Tune A samples. The difference in the measured $b$-jet shapes obtained using these two Monte Carlo samples is taken as a systematic uncertainty.\\
To gauge the effect on the unfolding procedures of any possible mis-modeling in the simulation of the detector response, the jet shapes are measured using the charged tracks inside the jet cone, instead of the calorimeter towers. The measured track jet shapes are unfolded back to hadron level using new correction functions obtained from MC simulation. The jet direction and transverse momentum remain unchanged with respect to the default scenario. All tracks with $p_T > 0.5$~GeV/$c$ that originate within a cone of 0.7 from the jet axis are considered for the measurement of these jet shapes. At detector level, the jet shapes defined by tracks tend to be narrower than those defined by calorimeter towers. The hadron level corrections will thus be different for the measurement using tracks than for the one using calorimeter towers The difference between the final hadron level jet shapes computed using this method and the default one is taken as a systematic uncertainty. \\
To investigate the accuracy of the simulation of the calorimeter response to low $p_T$ particles, the analysis is performed using only calorimeter towers with $p_T>0.5$~GeV/$c$ and the difference with respect to the nominal measurement (which uses calorimeter towers with $p_T > 0.1$~GeV/$c$) is taken as a systematic uncertainty and is found to be negligible.\\
A $3\%$ systematic uncertainty on the jet energy corrections is considered that combines the $3\%$ systematic uncertainty for inclusive jets~\cite{jet_corr} with the uncertainty on the $b$-jet fragmentation that is $0.6\%$ . A variation of $\pm15\%$ on the Missing $E_T$ significance is applied. The cut on the location of the primary vertex is varied by $\pm$5~cm around the nominal cut at 50~cm. These variations are all found to have only small effects on the final measurement. The dependence on the Monte Carlo modeling of the secondary vertex parameters was also investigated and found to be negligible.\\
 The total, statistical and systematic uncertainties are shown in Fig.~\ref{fig:total_sys} for each $p_T$ bin and $r$ bin. Also shown are the various contributions from the dominant effects.\\
The dominant sources of systematic uncertainties vary as a function of the $p_T$ bin. These are:\\
the difference in the $b$-jets shapes reconstructed using {\sc pythia} Tune A and {\sc herwig};\\
the difference in the $b$-jet shapes reconstructed from tracks instead of calorimeter towers, i.e. due to the detector simulation;\\
the $f_\mathrm{1b}$ variation from -0.2 to -0.5;\\
the $f_\mathrm{1c}$ variation by -0.2;\\
the jet energy scale.

%
\section{Results}\label{sec:results}
\begin{figure*}[h]
  \begin{center}
\ifprd
    \includegraphics[width=18.0cm,clip=]{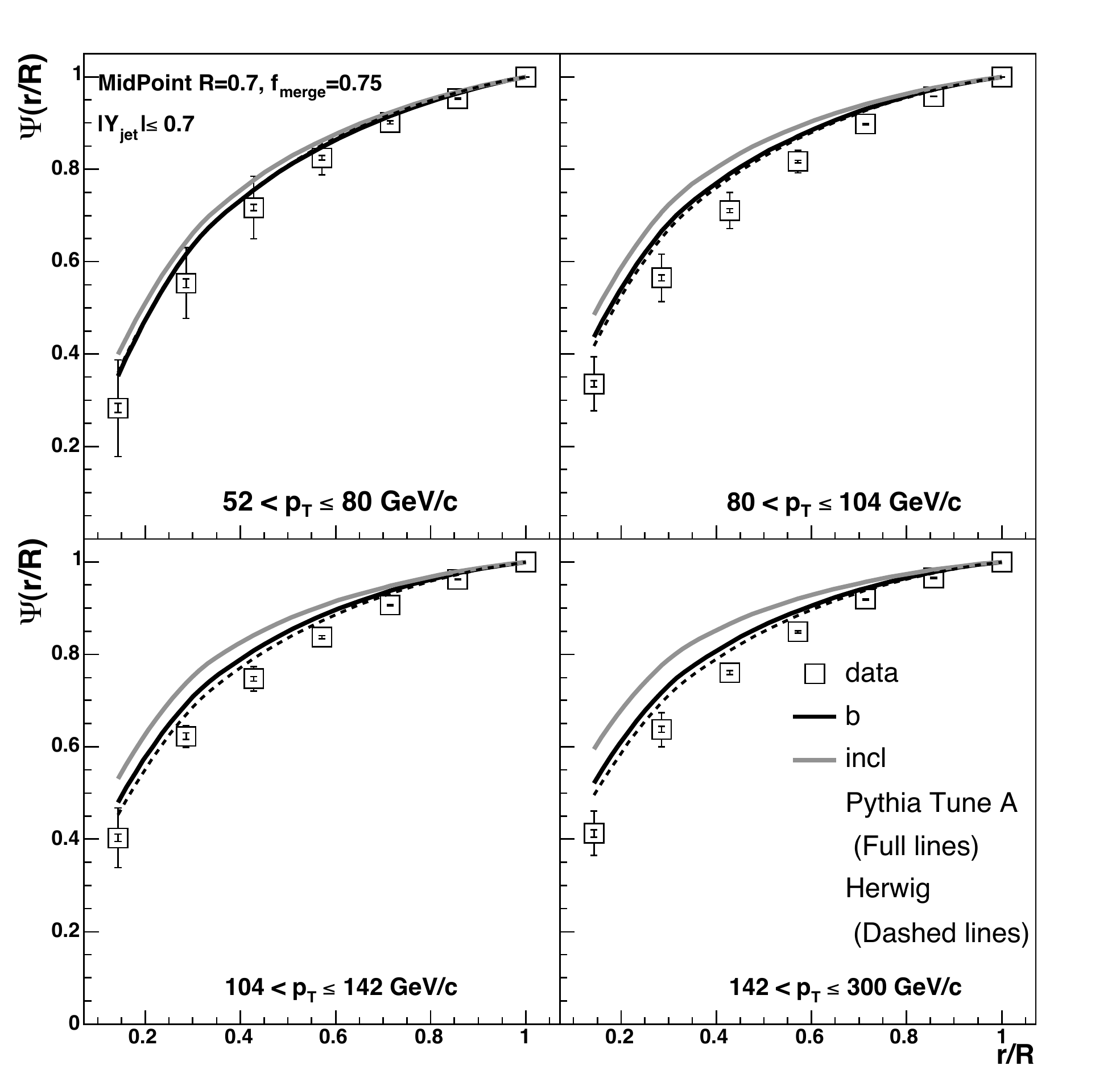}
 \else
    \includegraphics[width=15.0cm,clip=]{unfolded_b_incl.pdf}
\fi
    \caption{Measured integrated $b$-jet shapes for each of the four $p_T$ bins considered. The results are shown as black open squares where the error bars represent the statistical and total uncertainties. The statistical uncertainties are smaller than the squares. The results are compared to {\sc pythia} Tune A (full lines) and {\sc herwig} (dashed lines) predictions using the default $f_\mathrm{1b}$ fractions (black lines).  Also shown are the {\sc pythia} Tune A predictions for the inclusive jet shapes (grey lines).\label{fig:b_jet_shapes_final_a}}
  \end{center}
  \end{figure*}

\begin{figure*}[h]
  \begin{center}
       \ifprd
       \includegraphics[width=18.0cm,clip=]{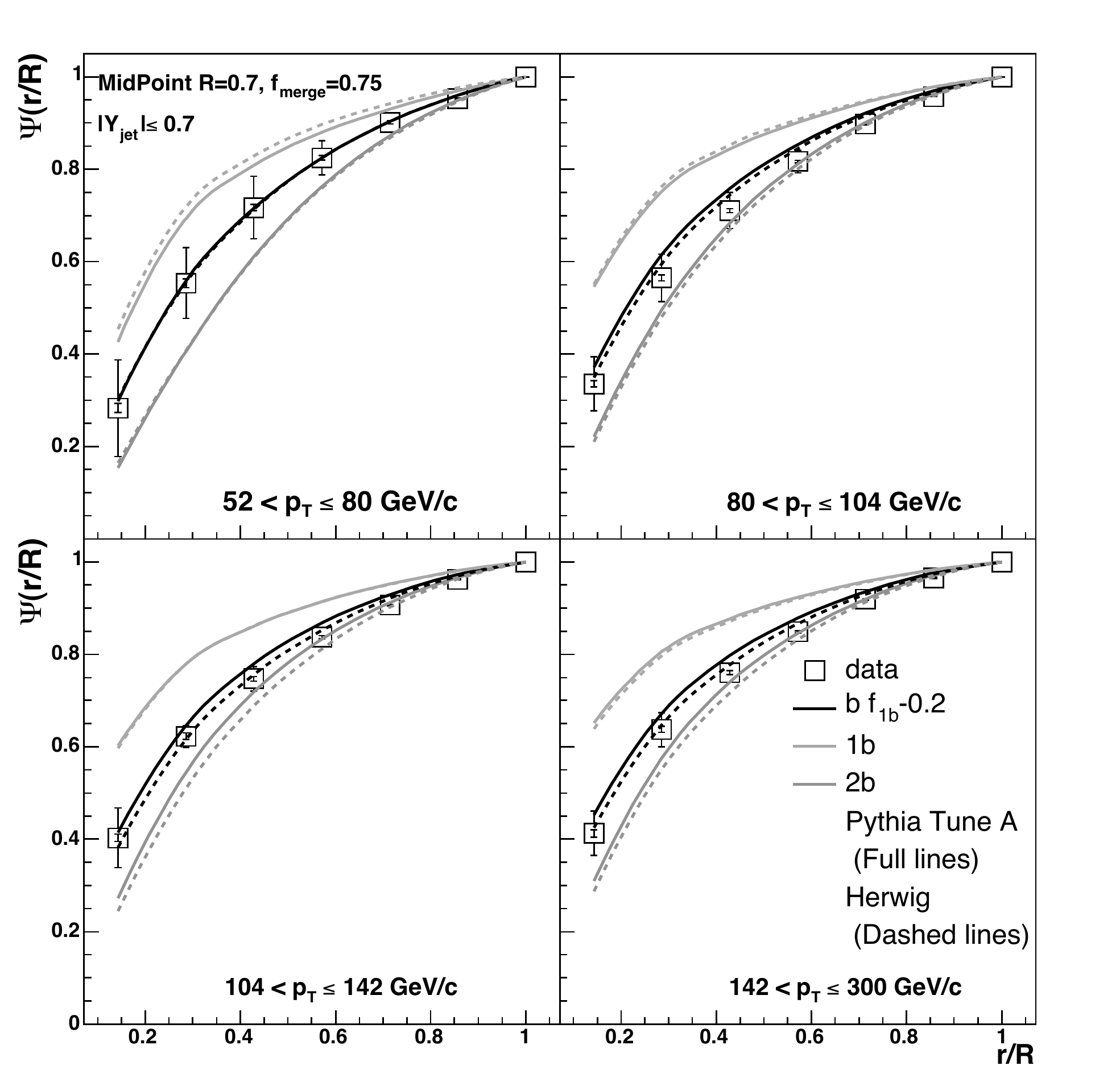}
\else
       \includegraphics[width=15.0cm,clip=]{unfolded_b_1b_2b.pdf}
\fi
    \caption{This plot shows the same data points as in Fig.~\ref{fig:b_jet_shapes_final_a}. The results are compared to {\sc pythia} Tune A (full lines) and {\sc herwig} (dashed lines) predictions if $f_\mathrm{1b}$ is decreased by 0.2 (black lines). Also shown are the {\sc pythia} Tune A and {\sc herwig} predictions for single and double $b$-quark jets (light, highest, and dark, lowest, grey lines, respectively).\label{fig:b_jet_shapes_final_b}}
  \end{center}
  \end{figure*}

\begin{figure*}[h]
  \begin{center}
  \ifprd
  \includegraphics[width=18.0cm,clip=]{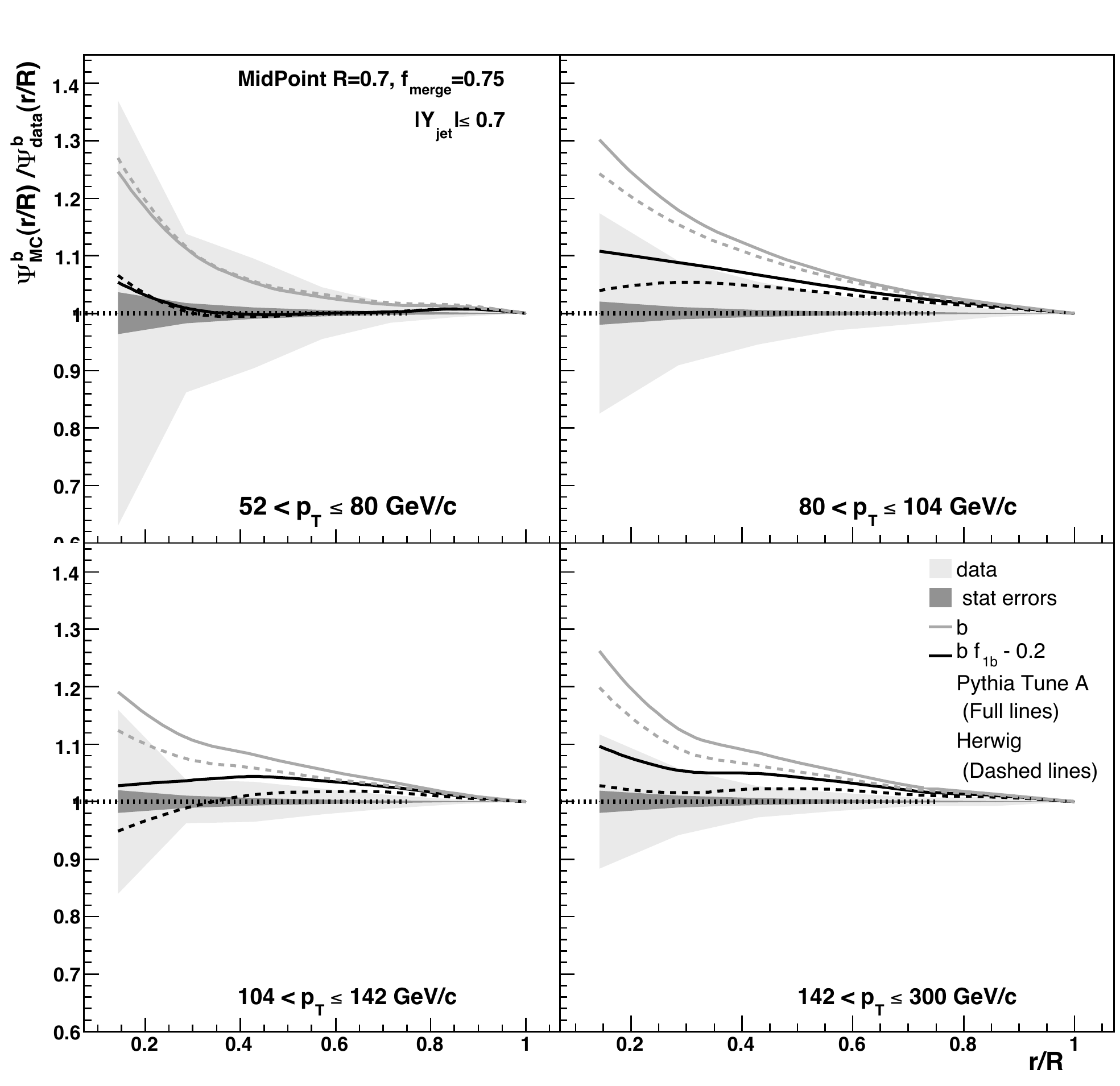}
  \else
  \includegraphics[width=15.0cm,clip=]{ratio_unfolded_b_1b_2b.pdf}
  \fi
    \caption{Ratios of the hadron level integrated $b$-jet shapes for various Monte Carlo predictions over the measured values. The light grey bands represent the total uncertainties on the measured $b$-jet shapes and the dark grey bands show the statistical uncertainties. {\sc pythia} Tune A and {\sc herwig} predictions using the default $f_\mathrm{1b}$ fractions are shown as grey lines (full lines for {\sc pythia} Tune A, dashed ones for {\sc herwig}). {\sc pythia} Tune A and {\sc herwig} predictions using $f_\mathrm{1b}$ fractions $0.2$ below the default values are also reported (black lines).
    \label{fig:shape_b_ratio}}
  \end{center}
  \end{figure*}

\begin{figure*}[h]
  \begin{center}
   \ifprd
    \includegraphics[width=18.0cm,clip=]{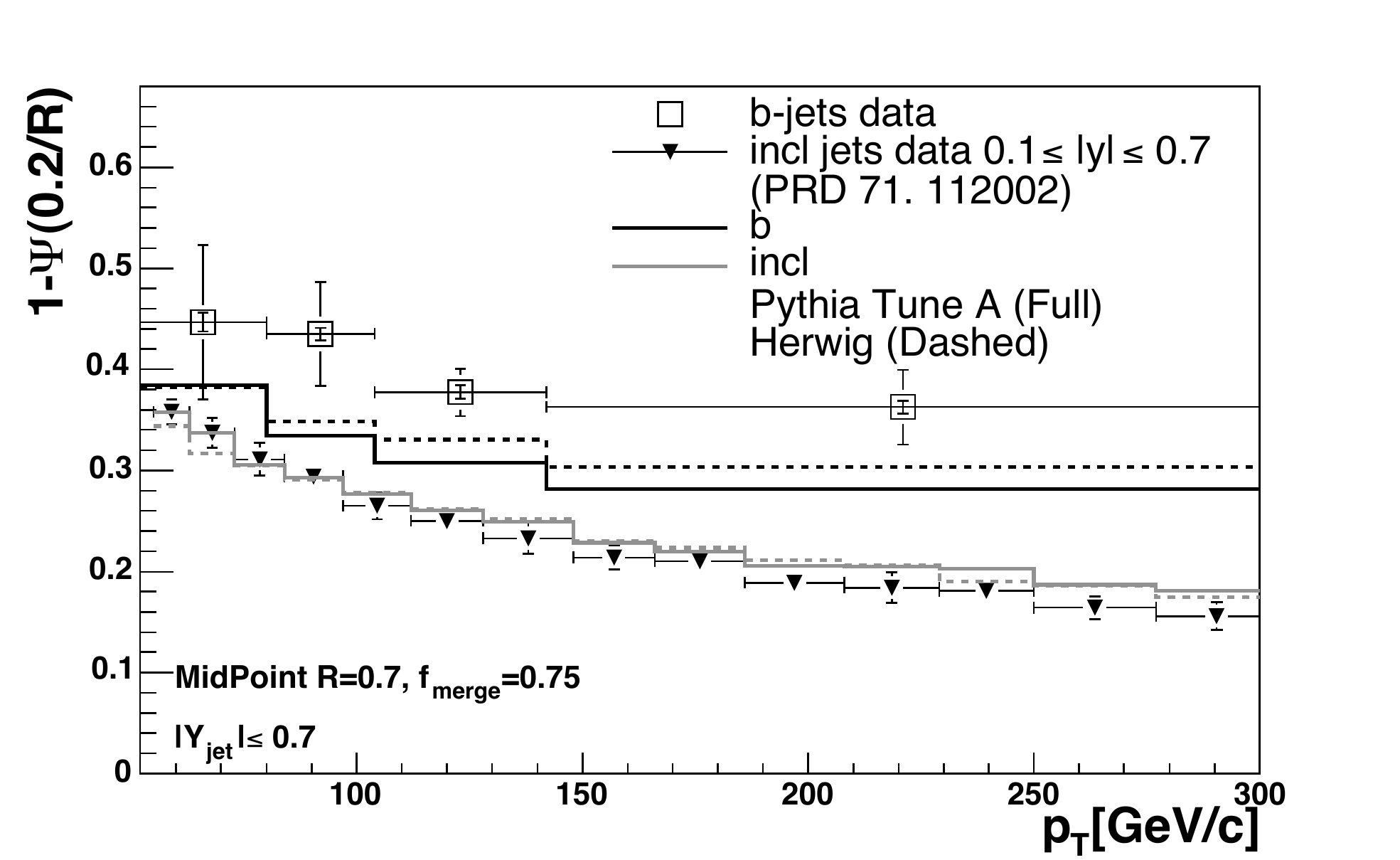}
    \includegraphics[width=18.0cm,clip=]{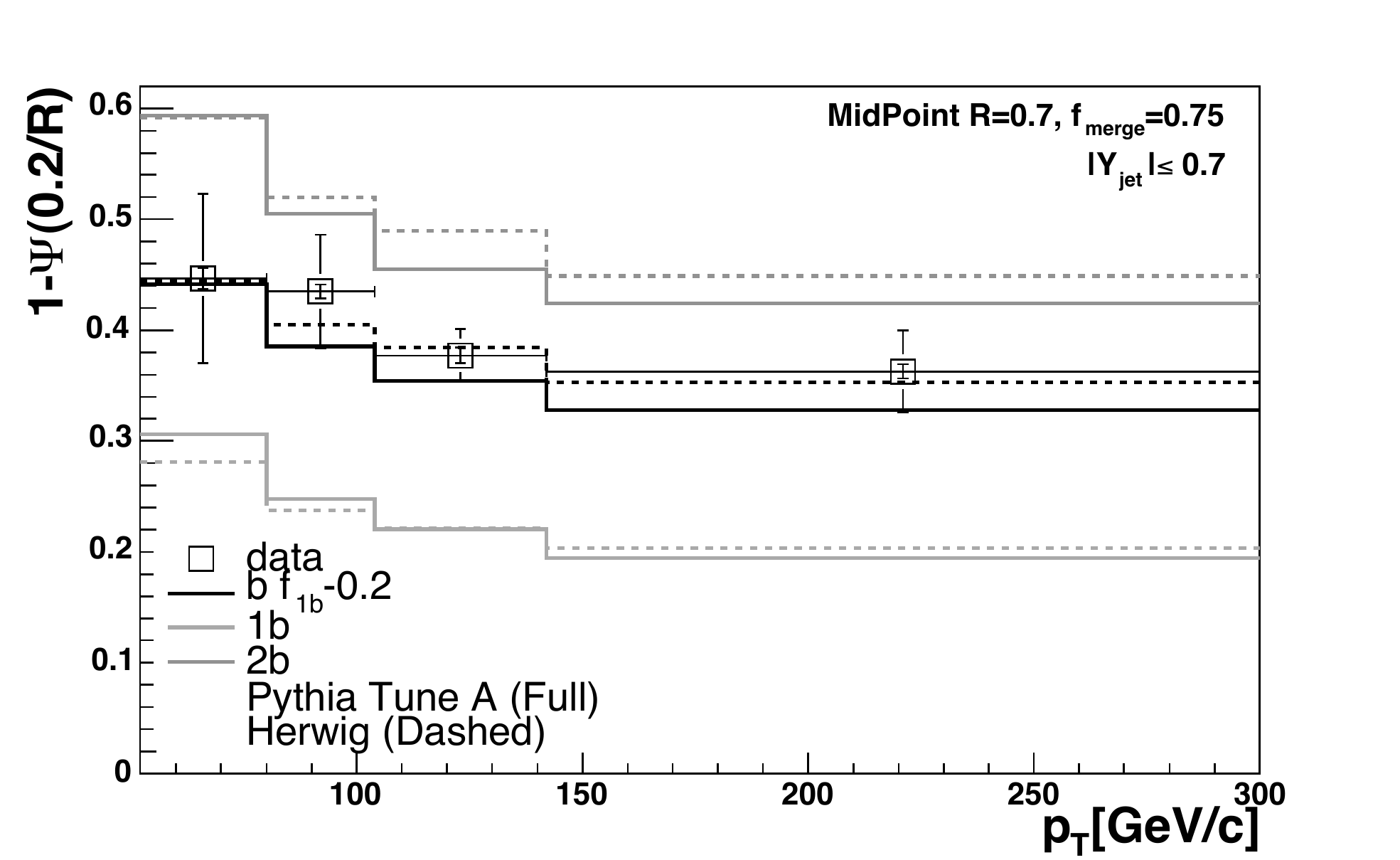}
    \else
    \includegraphics[width=10.0cm,clip=]{shape_b_incl_pt_02.pdf}
    \includegraphics[width=10.0cm,clip=]{shape_b_1b_2b_pt_02.pdf}    
    \fi
     \caption{Fractional $p_T$ outside a cone of radius $r= 0.2$ around the jet axis as a function of the $p_T$ of the jet. The results for $b$-jets are shown as black open squares and compared to different {\sc pythia} Tune A and {\sc herwig} predictions. The error bars on the plots represent the total and statistical uncertainties. The statistical uncertainties are smaller than the squares. (top) The results are compared to {\sc pythia} Tune A (full lines) and {\sc herwig} (dashed lines) predictions using the default $f_\mathrm{1b}$ fractions (black lines).  Also shown are the {\sc pythia} Tune A predictions for the inclusive jet shapes (grey lines) as well as the previously published inclusive jet shape results (triangles). (bottom) The measured values are shown along with the expected distributions if $f_\mathrm{1b}$ is decreased by 0.2 (black lines). Also shown are the {\sc pythia} Tune A and {\sc herwig} predictions for single and double $b$-quark jets (light, lowest, and dark, highest, grey lines, respectively). \label{fig:b_jet_shapes_pt}}
  \end{center}
  \end{figure*}

The final results are presented and discussed in this section. The measured integrated $b$-jet shapes are shown in Figs.~\ref{fig:b_jet_shapes_final_a} and~\ref{fig:b_jet_shapes_final_b} as black open squares. The statistical and total uncertainties on the measurements are shown; the statistical uncertainties are smaller than the points. The uncertainty on the measured jet shape is 0 at $r/R = 1$ because the integrated jet shape is defined to be exactly equal to 1 at this point. The first three bins in $r/R$ thus contain most of the information about the broadness of the jet shapes.\\
The results are compared to {\sc pythia} Tune A and {\sc herwig} predictions using the default $f_\mathrm{1b}$ fractions. Figure~\ref{fig:b_jet_shapes_final_a} also shows the {\sc pythia} Tune A predictions for the inclusive jet shapes. Reference~\cite{shapes_incl} shows good agreement between data and the MC predictions for inclusive jet shapes. This plot shows that despite relatively large systematic uncertainties, the measurements differ from the inclusive jet shape predictions, thus indicating that the jet shapes are sensitive to the presence of heavy flavor particles. Jets containing $b$-quarks appear to be broader than inclusive jets. No reasonable change in f-1b could
bring the data into agreement with the inclusive jet shapes. This plot also shows that the agreement between data and LO MC simulation is much better using a smaller fraction of jets that contain a single $b$-quark than the default value.\\
The {\sc pythia} Tune A and {\sc herwig} predictions, when the expected distributions if $f_\mathrm{1b}$ is decreased by 0.2, along with the predictions for single and double $b$-quark jets, are reported in Fig.~\ref{fig:b_jet_shapes_final_b}. Single $b$-quark jets are predicted to be narrower than inclusive $b$-jets; double $b$-quark jets are predicted to be broader. The measured average $b$-jet shapes are between these two curves, narrower than double $b$-quark jets but broader than single $b$-quark jets. It appears from these plots that a larger fraction of double $b$-quark jets than predicted by the LO MC simulation agrees better with the data. The MC method predictions for {\sc pythia} Tune A and {\sc herwig} are shown to be similar for all cases considered.\\
Figure~\ref{fig:shape_b_ratio} shows the ratio of the hadron level integrated jet shapes from Monte Carlo predictions over the measured values for each of the four jet $p_T$ bins. The light grey bands show the total uncertainty on the measurement, whereas the dark grey band shows the statistical uncertainty. From these plots it is clear that the agreement between data and MC simulation is improved in each $p_T$ bin by  decreasing the single $b$-quark fraction by 0.2.\\
Table~\ref{tab:results} reports the measured $b$-jet shapes in each of the four $p_T$ bins for each of the bins in $r$. The central values are shown along with the statistical and systematic uncertainties.\\
 \begin{table}[h]
 \ifprd
\begin{small}
\else
\begin{scriptsize}
\fi
\begin{center}
\begin{tabular}{cc} \hline\hline
  \multicolumn{2}{c}{$52 \leq p_T<80$~GeV/$c$} \\  
\hline
 $r/R$ & $\Psi^\mathrm{b}(r/R) \pm \sigma_\mathrm{stat} \pm \sigma_\mathrm{sys}$ \\
\hline 
$0.1/0.7\,\approx\, 0.14 $ &$ 0.283 \pm 0.010 \pm 0.105 $ \\ 
$0.2/0.7 \,\approx\,0.28$ &$ 0.553 \pm 0.010 \pm 0.076 $ \\ 
$0.3/0.7  \,\approx\, 0.42$ &$ 0.717 \pm 0.007 \pm 0.068 $ \\ 
$0.4/0.7  \,\approx\, 0.57$ &$ 0.825 \pm 0.005 \pm 0.037 $ \\ 
$0.5/0.7  \,\approx\, 0.71$ &$ 0.901 \pm 0.003 \pm 0.015 $ \\ 
$0.6/0.7  \,\approx\, 0.86$ &$ 0.953 \pm 0.002 \pm 0.006 $ \\ 
$0.7/0.7\,=\,1.00$ &$ 1.000 \pm 0.000 \pm 0.000 $ \\ 
\hline
\hline
 \multicolumn{2}{c}{$80 \leq p_T < 104$~GeV/$c$}\\
 \hline
 $r/R$ & $\Psi^\mathrm{b}(r/R) \pm \sigma_\mathrm{stat} \pm \sigma_\mathrm{sys}$ \\
 \hline
$0.1/0.7\,\approx\, 0.14$  &$ 0.336 \pm 0.007 \pm 0.059 $ \\ 
$0.2/0.7 \,\approx\,0.28$  &$ 0.565 \pm 0.006 \pm 0.051 $ \\ 
$0.3/0.7 \,\approx\, 0.42$&$ 0.710 \pm 0.004 \pm 0.039 $ \\ 
$0.4/0.7  \,\approx\, 0.57$&$ 0.817 \pm 0.003 \pm 0.024 $ \\ 
$0.5/0.7 \,\approx\, 0.71$&$ 0.898 \pm 0.002 \pm 0.017 $ \\ 
$0.6/0.7  \,\approx\, 0.86$ &$ 0.957 \pm 0.001 \pm 0.006 $ \\ 
$0.7/0.7 \,=\,1.00$&$ 1.000 \pm 0.000 \pm 0.000 $ \\ 
\hline
\hline
\multicolumn{2}{c}{$104 \leq p_T < 142$~GeV/$c$}\\
\hline
 $r/R$ & $\Psi^\mathrm{b}(r/R) \pm \sigma_\mathrm{stat} \pm \sigma_\mathrm{sys}$ \\
 \hline
$0.1/0.7\,\approx 0.14$  &$ 0.403 \pm 0.008 \pm 0.064 $ \\ 
$0.2/0.7 \,\approx\,0.28$  &$ 0.623 \pm 0.007 \pm 0.024 $ \\ 
$0.3/0.7 \,\approx\, 0.42$&$ 0.747 \pm 0.005 \pm 0.026 $ \\ 
$0.4/0.7 \,\approx\, 0.57$&$ 0.837 \pm 0.003 \pm 0.019 $ \\ 
$0.5/0.7 \,\approx\, 0.71$&$ 0.906 \pm 0.002 \pm 0.010 $ \\ 
$0.6/0.7  \,\approx\, 0.86$ &$ 0.963 \pm 0.001 \pm 0.004 $ \\ 
$0.7/0.7 \,=\,1.00$&$ 1.000 \pm 0.000 \pm 0.000 $ \\ 
\hline
\hline
\multicolumn{2}{c}{$142 \leq p_T < 300$~GeV/$c$}\\
 \hline
 $r/R$ & $\Psi^\mathrm{b}(r/R) \pm \sigma_\mathrm{stat} \pm \sigma_\mathrm{sys}$ \\
 \hline
$0.1/0.7\,\approx 0.14 $ &$ 0.413 \pm 0.008 \pm 0.048 $ \\ 
$0.2/0.7 \,\approx\,0.28$  &$ 0.637 \pm 0.006 \pm 0.037 $ \\ 
$0.3/0.7 \,\approx\, 0.42$&$ 0.760 \pm 0.005 \pm 0.020 $ \\ 
$0.4/0.7  \,\approx\, 0.57$&$ 0.849 \pm 0.003 \pm 0.013 $ \\ 
$0.5/0.7 \,\approx\, 0.71$&$ 0.919 \pm 0.002 \pm 0.007 $ \\ 
$0.6/0.7  \,\approx\, 0.86$ &$ 0.966 \pm 0.001 \pm 0.008 $ \\ 
$0.7/0.7 \,=\,1.00$&$ 1.000 \pm 0.000 \pm 0.000 $ \\ 
\hline
\end{tabular}
\caption{Integrated jet shapes for $b$-jets. The central value along with the statistical and systematic uncertainty for each $p_T$ and $r$ bin are shown.}\label{tab:results}
\end{center}
\ifprd
\end{small}
\else
\end{scriptsize}
\fi
\end{table}
 Another way of looking at these results is to plot the fractional $p_T$ outside a cone of fixed radius
$r$ as a function of the $p_T$ of the jets. This gives an idea of the change in width of the jets as the transverse momentum increases. Jets of a particular flavor are expected to become narrower as the $p_T$ increases, mainly due to the running of the strong coupling constant, $\alpha_s$. Figure~\ref{fig:b_jet_shapes_pt} shows the evolution with jet $p_T$ of the measured fractional $p_T$ outside a cone of fixed radius r=0.2. The results are compared to {\sc pythia} Tune A and {\sc herwig} predictions using the default $f_\mathrm{1b}$ fractions as well as the expected distributions if $f_\mathrm{1b}$ is decreased by 0.2. The {\sc pythia} Tune A predictions for the inclusive jet shapes as well as the previously published inclusive jet shape results are shown in the top plot. The rapidity region considered for the inclusive jet shape measurement does not include the center-most rapidity region ($|y|<0.1$). The inclusion or not of this region was found in this analysis not to change the value of the predictions or the measured values. This figure indicates that the evolution of the jet shape with $p_T$ appears to be somewhat flatter for $b$-jets than for inclusive jets and confirms that the measured b-jet shapes are different from those measured for inclusive jets.\\
The {\sc pythia} Tune A and {\sc herwig} predictions for single and double $b$-quark jets are included on the bottom plot. This figure shows that as the $p_T$ of the jet increases, the difference between the {\sc pythia} Tune A and {\sc herwig} predictions increases. {\sc herwig} predicts a slightly flatter evolution with $p_T$ than {\sc pythia} Tune A.

\section{Conclusions}\label{sec:conclusions}
We have reported on a measurement of the b-jet shape at the Tevatron collider. Despite the considerable uncertainties of Monte Carlo simulations of this non-perturbative process, we show convincing evidence that b-jets are broader than inclusive ones. This confirms that the jet shape is sensitive to the heavy flavor content.\\
The measured $b$-jet shapes are significantly broader than expected from both {\sc pythia} Tune A and {\sc herwig}. One possible interpretation is that this effect is coming from an underestimation in LO MC simulation of the fraction of $b$-jets originating from gluon splitting. NLO calculations predict a significantly higher rate of $b$-jets that contain more than one $b$-quark inside the jet cone than the LO Monte Carlos. Decreasing the relative fraction of single $b$-quark jets, i.e. increasing the double $b$-quark jet fraction, by 0.2 in {\sc pythia} Tune A and {\sc herwig} leads to a better description of the measured $b$-jet shapes. This decrease is qualitatively consistent with the NLO predictions with a small factorisation and renormalisation scale, $\mu = \frac{\mu_0}{2}$, where $\mu_0=\sqrt{p_T^2+m_b^2}$. These findings are consistent with a number of other analyses that investigated the azimuthal correlations in $b\bar{b}$ production.
%
%
\ifprd
\begin{acknowledgments}
We thank the Fermilab staff and the technical staffs of the participating institutions for their vital contributions. This work was supported by the U.S. Department of Energy and National Science Foundation; the Italian Istituto Nazionale di Fisica Nucleare; the Ministry of Education, Culture, Sports, Science and Technology of Japan; the Natural Sciences and Engineering Research Council of Canada; the National Science Council of the Republic of China; the Swiss National Science Foundation; the A.P. Sloan Foundation; the Bundesministerium f\"ur Bildung und Forschung, Germany; the Korean Science and Engineering Foundation and the Korean Research Foundation; the Science and Technology Facilities Council and the Royal Society, UK; the Institut National de Physique Nucleaire et Physique des Particules/CNRS; the Russian Foundation for Basic Research; the Ministerio de Educaci\'{o}n y Ciencia and Programa Consolider-Ingenio 2010, Spain; the Slovak R\&D Agency; and the Academy of Finland.
\end{acknowledgments}
\fi

\newpage

\begin{thebibliography}{99}
\bibitem{foot:eta}
The transverse momentum, $p_T$, is defined as the momentum transverse to the beam axis; 
The transverse energy $E_T$, is defined as the energy multiplied by the sine of the angle with respect to the beam axis; The pseudorapidity is defined as $\eta = - \ln{(\tan(\theta/2))}$, where the polar angle $\theta$ is taken with respect to the proton beam direction. The rapidity is defined as $y = \frac{1}{2} \ln{(\frac{E+p_z}{E-p_z})}$ where $E$ denotes the energy and $p_z$ is the component of the momentum along the beam direction. $\phi$ is measured in the plane orthogonal to the beam direction.
\bibitem{prescales}
In order to reduce online trigger rates, different prescale factors are applied to trigger paths. For a given trigger path, a prescale of 1000 would indicate that only one in every 1000 events satisfying the trigger requirements is selected.
\bibitem{foot:d0_sig}
The significance of the distance of closest approach, $d_0$, is defined as $\frac{d_0}{\sigma(d_0)}$, where $\sigma(d_0)$ is the uncertainty on this value.
\bibitem{foot:lxy}
The significance of the 2-dimensional projection of the secondary vertex displacement is defined as $\frac{L_{xy}}{\sigma(L_{xy})}$, where $L_{xy}$ is the 2-dimensional projection and, $\sigma(L_{xy})$ is the uncertainty on this value.
\end{thebibliography}
\end{document}